# The Emergence of China as a Leading Nation in Science

*Research Policy* (forthcoming)

Ping Zhou[a] * & Loet Leydesdorff [b]

[a] *Information Research and Analysis Center. Institute of Scientific and Technical Information of China. 15 Fuxing Road. Beijing 100038. P. R. China;*
*zhoup@istic.ac.cn*

[b] *Amsterdam School of Communications Research (ASCoR). University of Amsterdam Kloveniersburgwal 48. 1012 CX Amsterdam. The Netherlands;*
*loet@leydesdorff.net ; http://www.leydesdorff.net*

**Abstract**

China has become the fifth leading nation in terms of its share of the world's scientific publications. The citation rate of papers with a Chinese address for the corresponding author also exhibits exponential growth. More specifically, China has become a major player in critical technologies like nanotechnology. Although it is difficult to delineate nanoscience and nanotechnology, we show that China has recently achieved a position second only to that of the USA. Funding for R&D has been growing exponentially, but since 1997 even more in terms of business expenditure than in terms of government expenditure. It seems that the Chinese government has effectively used the public-sector research potential to boost the knowledge-based economy of the country. Thus, China may be achieving the ("Lisbon") objectives of the transition to a knowledge-based economy more broadly and rapidly than its western counterparts. Because of the sustained increase in Chinese government funding and the virtually unlimited reservoir of highly-skilled human resources, one may expect a continuation of this growth pattern in the near future.

* Temporary address of the corresponding author: ASCoR, Kloveniersburgwal 48, 1012 CX Amsterdam, The Netherlands; tel.: +31 20 5256174; fax: +31-20- 525 3681; email: P.Zhou@uva.nl.

## 1. Introduction

In recent years, China's economy has been growing fast. The average annual GDP growth rate was 9.9% during the period 1992-2001 (National Bureau of Statistics of China, 2002). This extraordinary growth is happening in a context in which some other countries are experiencing stagnation and/or recession. Such economic growth can be expected to have positive effects on China's scientific research and development (R&D) because, for example, more funding can be made available for R&D. Since the relation between knowledge-based innovations and the economy is interdependent and mutually enhancing (Foray, 2004), the growth of the scientific and technological capacities of China can be expected to reinforce its economic development.

In the transition to a knowledge-based economy, R&D expenditure has been considered as an important indicator for evaluating a country's investment in its knowledge base. The European Summit of 2000 in Lisbon, for example, agreed to strive for a ratio of 3% Gross Expenditure in R&D (GERD) over GDP in 2010 (European Commission, 2000, 2005). The ratio of GERD/GDP for China has been increasing exponentially during the last decade despite the spectacular increase in the denominator (GDP). In 2003, the GERD/GDP-ratio for China was 1.31%, while only five years ago this ratio was 0.70% (China Science and Technology Statistics Data Book, 2004).

Another factor closely related to the emergence of Chinese science is the return of overseas scholars. China's rapid and sustained economic development has motivated an increasing number of overseas scholars to return (Wang & Zheng, 2005). In order to encourage overseas Chinese scholars to join the construction of the Chinese knowledge-based economy, Chinese governments at various levels have developed policies favourable to the return of emigrants. As a result, 81% of the members of the Chinese Academy of Sciences and 54% of those of the Chinese Academy of Engineering are returned overseas scholars (Xing, 2004). These returned overseas scholars play important roles in China's economic and scientific development.

Is the output of these efforts keeping pace with the input? If so, how is the impact of publications in terms of citations developing? (Jin & Rousseau, 2004). In a paper published in *Nature*, the Chief Scientific Adviser to the UK Government, Sir David A. King, compared certain major countries including the USA, the EU countries, and China (King, 2004). Among other indicators, the author suggested a relationship between wealth intensity (GDP per person) and citation intensity (citations per paper). On this indicator for the impact of scientific performance, China was at the very bottom, just above India and Iran. In a reaction to King's (2004) paper, however, Ronald Kostoff—a scientometrician working for the US Navy—called the report "misleading" because it underestimated the emerging role of China in critical technologies like nanotechnology. Kostoff (2004) claimed that if a composed indicator is used, China could be shown to have surpassed the United States during the first eight months of 2004 in terms of research output in this field. On Kostoff's indicator the UK figures only seventh after China, the USA, Japan, Germany, France, and South Korea.

These conflicting views inspired us to analyze China's performance in R&D using scientometric indicators (Zhou & Leydesdorff, 2004; Leydesdorff & Zhou, 2005). In this study, we extend the scientometric analysis with input statistics as provided by the OECD, a more detailed analysis of China's performance in nanotechnology, and an analysis of the Chinese publication system in terms of international and domestic journals. Thus, we are able to provide a more integrated picture of the knowledge base of the Chinese system. In the final sections, we draw some conclusions and suggest normative implications for the further development of China's research potential.

## 2. Methods and Materials

The scientometric analysis is based on using the various versions of the *Science Citation Index*. We use indicators like total publications, world share of publications, total citation rates, percentage of world share of citations, as well as the top one percent of most highly cited papers in order to measure scientific output. The analysis focuses on the six major countries (the USA, Japan, UK, Germany, France, and China), and we added the EU-15 and the EU-25 because this provides an additional perspective at the global level. We also included South Korea because this comparison may teach us something about the differences in the dynamics between Asian versus other OECD countries. (Korea has been a member of the OECD since 1996.)

For the input indicators, we used the OECD's *Main Science and Technology Statistics* published online and in print (OECD, 2004). One should note that the normalization of the Chinese currency in terms of its equivalent purchasing power parity in U.S. dollars has remained a subject of some discussion (Davies, 2003; Shi, 2004). The online data was retrieved in the period between November 24, 2004 and January 24, 2005.

For the delineation of the field of nanoscience and nanotechnology, we use statistical techniques which were developed in other contexts (Leydesdorff & Cozzens, 1993; Leydesdorff, 2004). These techniques are applied to aggregated journal-journal citation relations as provided by the *Journal Citation Reports* of the *Science Citation Index 2003*. Since "nanotechnology" as a field of science is highly interdisciplinary, we experiment with different delineations of the field in terms of relevant journals. In addition to core-journals, we distinguish a nano-relevant set which forms the citation environment of the core journals and thus may provide the seedbed for further developments in this field. The performance data of nations in these limited sets can be compared with the performance indicators over the file of the *Science Citation Index*.

The Web-of-Science installation of the *Science Citation Index* allows for the measurements including the most recent year (2004), but there are some limitations on the retrieval. The system does not provide an exact number when the recall is larger than 100,000, and the download for each save is limited to 500. In order to solve the first problem, we separated the recalls that are larger than 100,000 into several smaller segments, then searched the results for each segment, and recombined them using Boolean algebra for the necessary corrections (because of international coauthorships). In the case of the EU-15 and EU-25, the correction for international coauthorships is not a *sine cura*. The number of Boolean operators increases rapidly with the number of sets to be combined.[1] At the level of each unit of analysis (country or set of countries) we use integer counting. Publications with an address in Hong Kong were merged with the data for China both before and after 1997 (the year when Hong Kong was returned to China) in order to prevent trend breaches.

In accordance with current practice in scientometrics (Braun *et al*., 1991), we have limited the analysis to articles, reviews, letters, and notes. For mapping the citation patterns of Chinese domestic journals, we used the same routines as for the nano-technology journals (Leydesdorff & Jin, 2005). These routines enable us to zoom into local structures by choosing an entrance journal for the analysis and then to visualize the relevant environments. The algorithm of Kamada-Kawai (1989) will be used for the visualizations. All representations are based on using the cosine among the citation vectors of journals as a measure for similarity (in the citing and cited dimensions, respectively); cosine values smaller than 0.2 are suppressed and the line thickness varies with the value of the similarity measure.

**3. Results**

In order to provide a clear picture of China's research performance, this section is organized in four subsections. Each subsection specifically focuses on one topic. First, we discuss the results related to China's general performance in terms of publications and citations, respectively, in sections one and two. Publications and citations are two key indicators for evaluating research output. Data about nanotechnology follows in the third section. We compare China with major countries like the USA, the UK, Germany, France, Japan, the EU-15, and the EU-25, and we include South Korea in order to assess whether the effects which we found were specific to China or applicable more generally to Asian nations. In the final section of this part, we compare the relations between input and output indicators for these (sets of) countries.

[1] Let us call the search results of four subsets A, B, C, and D, respectively. Any two subsets can be combined using: $R = (A+B) - \cap AB$. For three subsets $R = (A+B+C) - ( \cap AB+\cap AC+\cap BC) + \cap ABC$. Adding a fourth segment D to R then requires the following calculations: $R' = (R+ D) - (\cap AD+\cap BD+\cap CD) + ( \cap ABD+\cap ACD+\cap BCD) - \cap ABCD$.

### 3.1 Publications with a Chinese address

Jin & Rousseau (2004) already signalled the exponentional growth in scientific publications with a Chinese address. In 1999, China was in the tenth position (China Science and Statistics Data Book, 2000). Five years later (2004) China had become the fifth largest country in terms of scientific publications, after the USA, Japan, the UK, and Germany, respectively. However, if we look at values for the EU-15 and the EU-25 on this indicator, the number of scientific publications of the EU-15 is 15% higher than that of the USA; and for the EU-25 the number is even more than 23% higher (Figure 1).

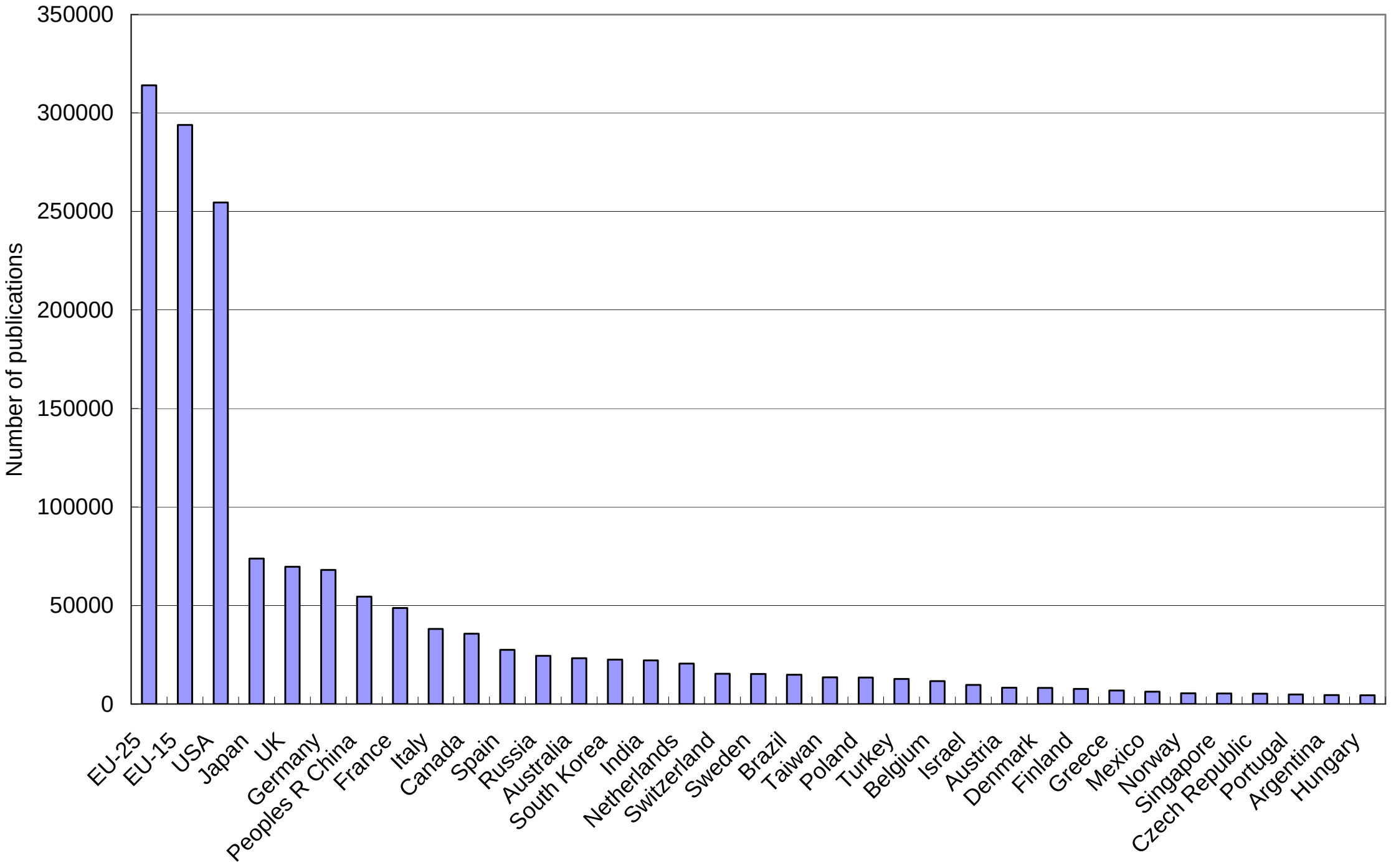

**Figure 1:** Ranks of countries in 2004 in terms of number of publications (articles, notes, reviews, and letters; *SCI Web-of-Science* 2004; data collected on 15 January 2005).

In order to show the historical development during the last decade, we collected data for the seven above-mentioned countries, the EU-15, and the EU-25 for the last ten years (Figure 2). China is the only country of which the percentage share shows exponential growth, while South Korea's growth trend was significant as well. However, Korea's increase trend is linear instead of exponential. Japan, the UK, and France were relatively stable; Germany's output showed an increase during the period 1995-1998 as an effect of the unification (Leydesdorff, 2000). The world share of publications of the EU-15 (articles, reviews, letters, and notes) surpassed that of the USA in 1994 (cf. King, 2004). The world share of publications of the EU-15 has been approximately 5% higher than that of the USA since 1998. The trend line of the EU-25 is similar to that of the EU-15. The USA's share decreased from 1995 to 2000, but the indicator is relatively stable thereafter.

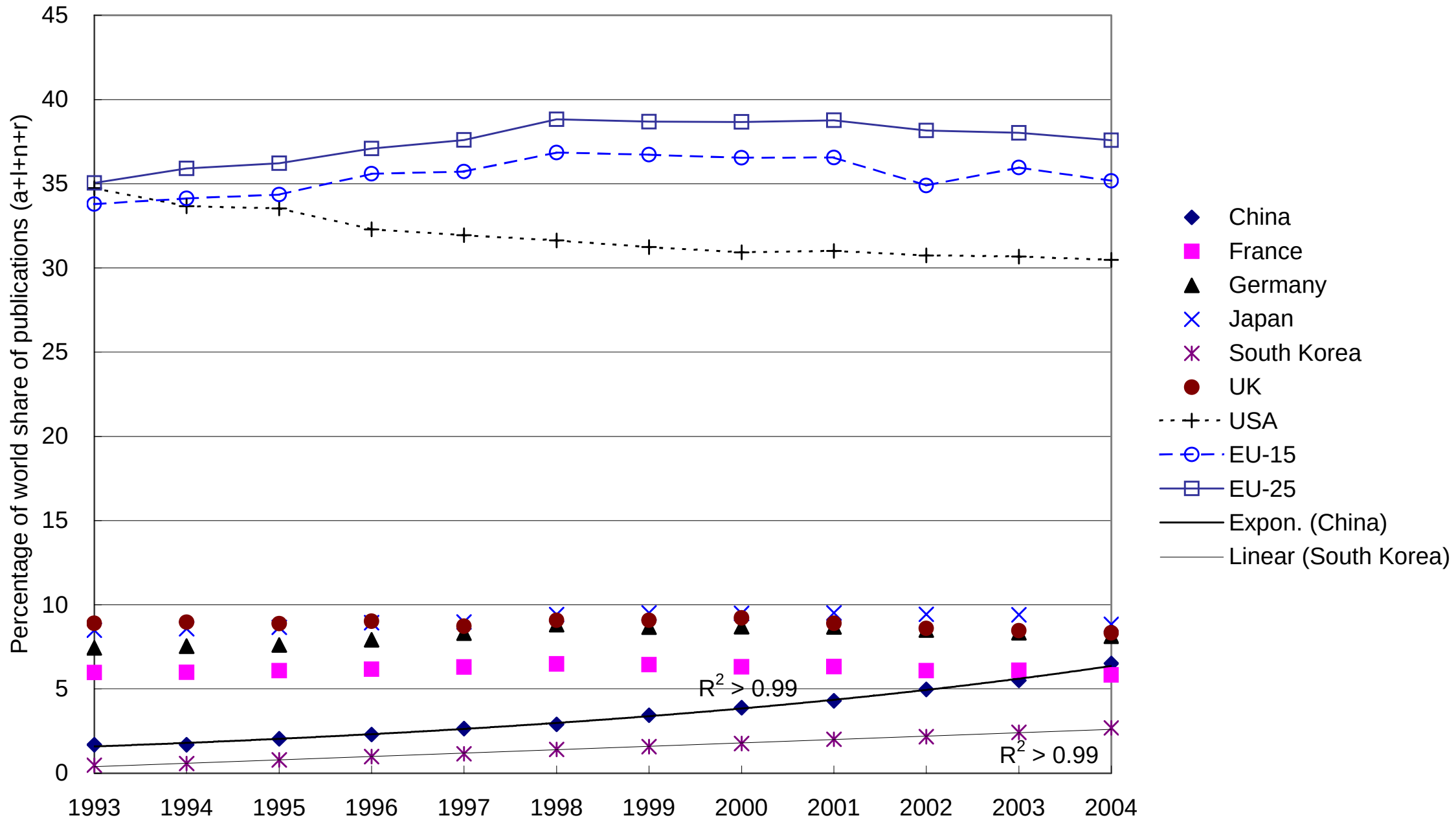

**Figure 2:** Percentage of world share in terms of articles, reviews, letters, and notes (*SCI Web-of-Science*, 1993-2004; data as of 29 January 2005).

Table 1 provides the percentages of world share in tabular format using the types of publications deemed most relevant for this assessment (articles, reviews, letters, and notes). From this table one can see in greater detail how the differences between countries have developed. The resulting picture is surprisingly dynamic. For example, Japan has surpassed the UK by taking the second position since 1997. China has been the only country to have its world share increase by more than one percent within a single year (2003-2004). Korea's sustained increase is also remarkable.

**Table 1:** Percentage of world share of the seven countries under study, EU-15 and EU-25 (*Web of Science* data).

| | China | France | FRG | Japan | Korea | UK | USA | EU-15 | EU-25 |
|---|---|---|---|---|---|---|---|---|---|
| *1993* | 1.69 | 5.98 | 7.45 | 8.49 | 0.48 | 8.89 | 34.73 | 33.78 | 35.04 |
| *1994* | 1.70 | 5.99 | 7.54 | 8.57 | 0.58 | 8.97 | 33.66 | 34.12 | 35.90 |
| *1995* | 2.05 | 6.09 | 7.62 | 8.65 | 0.79 | 8.88 | 33.54 | 34.36 | 36.21 |
| *1996* | 2.31 | 6.18 | 7.93 | 8.94 | 0.99 | 9.02 | 32.29 | 35.59 | 37.08 |
| *1997* | 2.66 | 6.31 | 8.32 | 8.98 | 1.16 | 8.73 | 31.94 | 35.72 | 37.60 |
| *1998* | 2.90 | 6.48 | 8.82 | 9.42 | 1.41 | 9.08 | 31.63 | 36.85 | 38.82 |
| *1999* | 3.44 | 6.44 | 8.67 | 9.52 | 1.58 | 9.08 | 31.24 | 36.72 | 38.68 |
| *2000* | 3.89 | 6.31 | 8.69 | 9.49 | 1.76 | 9.22 | 30.93 | 36.55 | 38.67 |
| *2001* | 4.30 | 6.33 | 8.68 | 9.52 | 2.01 | 8.90 | 31.01 | 36.55 | 38.77 |
| *2002* | 4.98 | 6.10 | 8.50 | 9.43 | 2.17 | 8.60 | 30.75 | 34.89 | 38.16 |
| *2003* | 5.51 | 6.10 | 8.35 | 9.40 | 2.43 | 8.46 | 30.68 | 35.96 | 38.02 |
| *2004* | 6.52 | 5.84 | 8.14 | 8.84 | 2.70 | 8.33 | 30.48 | 35.18 | 37.59 |

In general, approximately 35% of world publications are from the EU countries, while approximately 30% is from the USA during the last three years. The Japanese share of publications wanders around 9%.

## 3.2 Citations to Chinese papers and Chinese journals

### *3.2.1 Citations to Chinese papers*

China's total citation rate is still low when compared with citation rates for other nations (Jin & Rousseau, 2003, 2005). However, this indicator has also been on the increase at an exponential rate during the last decade (ISTIC, 2003 and 2004). Figure 3 provides the number of citations in each year for publications with a Chinese address for the first author during the preceding ten years. (The first author is often the corresponding author.) In other words, a ten-year citation window is used on the set of articles, reviews, and letters in each year, including internationally coauthored publications, but only insofar as the *first* author has a Chinese address. The data are total citations without excluding self citations.

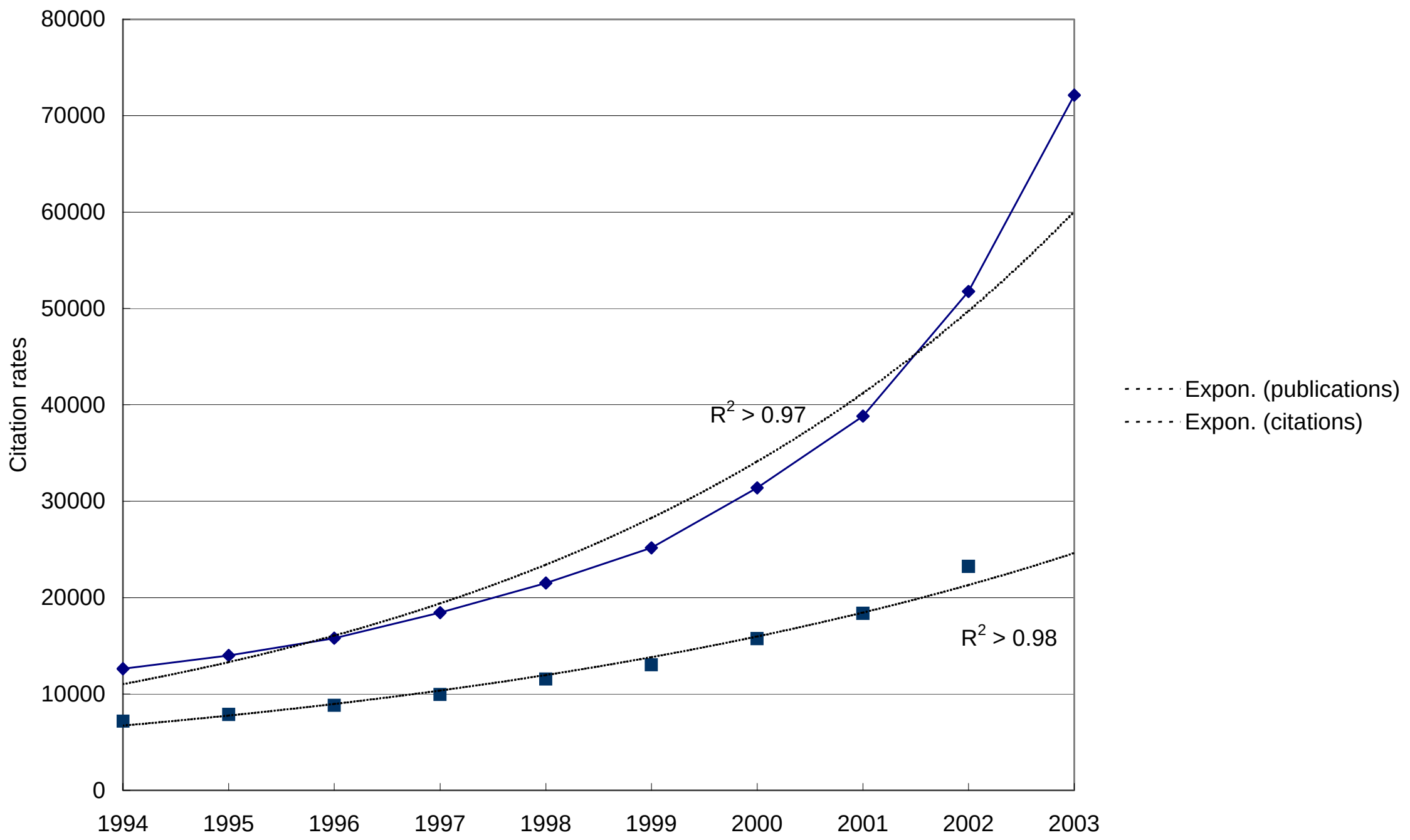

**Figure 3:** Citations to papers with a Chinese address using a ten-year citation window (Source: ISTIC, 2003 and 2004).

The figure shows that the increase in the citation rates is even *above* the best exponential fit of the curve during the last two years. This momentum of the Chinese publication system in terms of gaining citations can also be made visible by using the increase of the percentage of world share of citations provided by King (2004) in Table 1 of his paper. King's results can be used for the validation because they were derived from the same source as ours (Evidence, 2003).

King (2004) used his data for comparing among the nations at the world systems level, but one can use the same data for the analysis of growth rates between the two periods involved (Leydesdorff & Zhou, 2005). In Figure 4, based on Table 1 of King's paper, China is the fourth country in terms of the growth of its percentage of world share in citations during the period of 1997-2001 compared to the period 1993-1997. Other countries like Singapore and South Korea also show a spectacular increase in their citation rates when these two periods are compared at the level of individual nations. (The remarkable increase for Iran has probably to be explained in terms of the opening up of this country towards the international community during this period, but the underlying dynamics may be very different from those in China and Korea.

However, we refrain from analyzing the data for Iran because we don't have sufficient contextual knowledge about this country.)

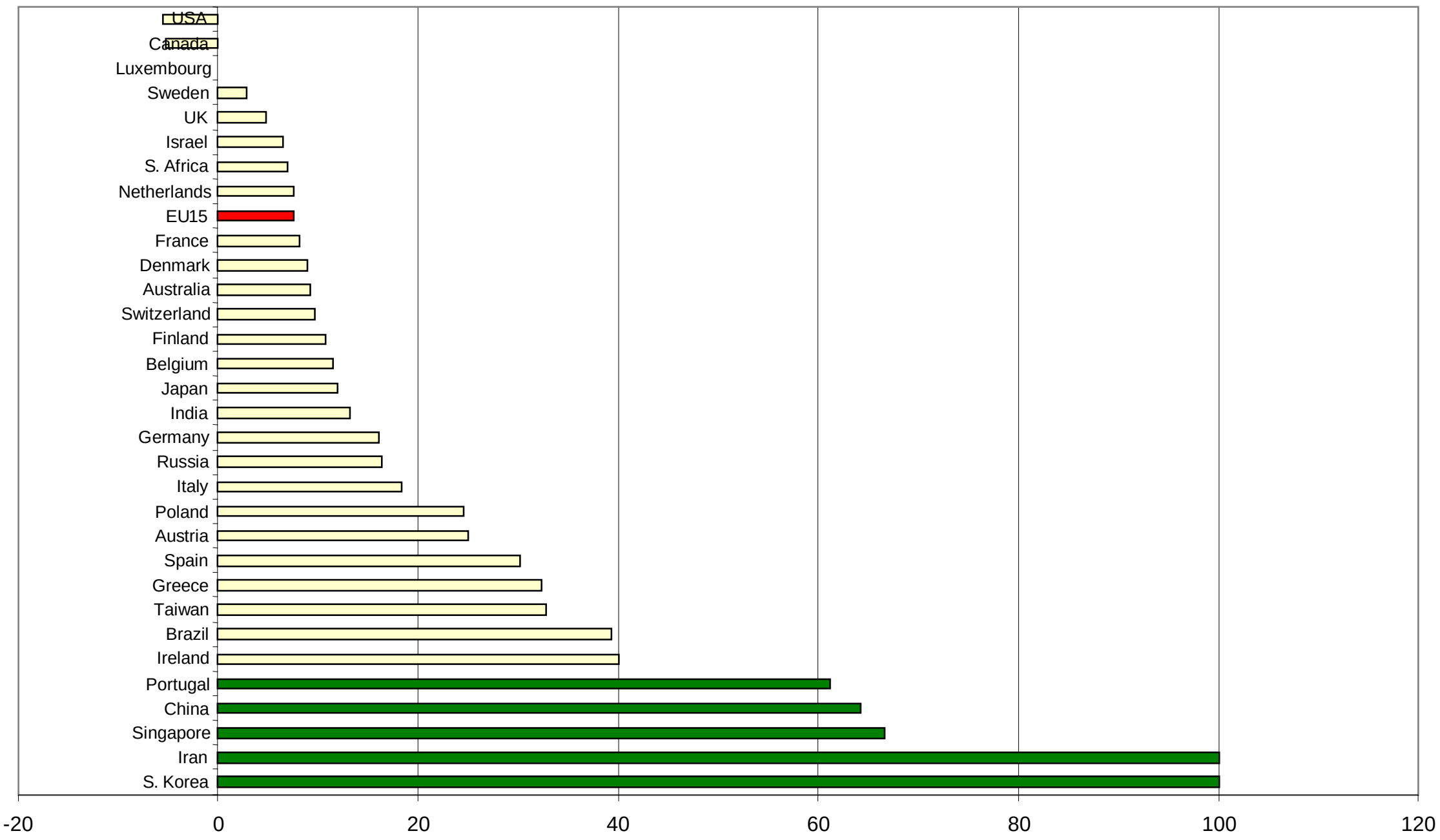

**Figure 4:** Increase in percentage of world share of citations. (Source: King, 2004, at p. 312).

King (2004) considered the indicator of the top one percent of the most highly cited papers as the most important measure of a country's influence in science. Using his data, we plotted the percentage change of a country's highly cited papers in Figure 5. (See for a discussion of King's metrics: Braun et al., 2005; Evidence, 2003). China's contribution increased on this indicator as well, although the absolute numbers are still low (0.22% and 0.33% for the two periods 1993-1997 and 1997-2001, respectively). Among the countries with an increase in the percentage of highly cited publications, China ranked sixth, with an increase almost similar to that of South Korea in the fifth position (45.9%). Note that English is native language in countries with an even higher increase (Ireland, India, South Africa, and Singapore). In other words, authors in these countries have a language advantage.

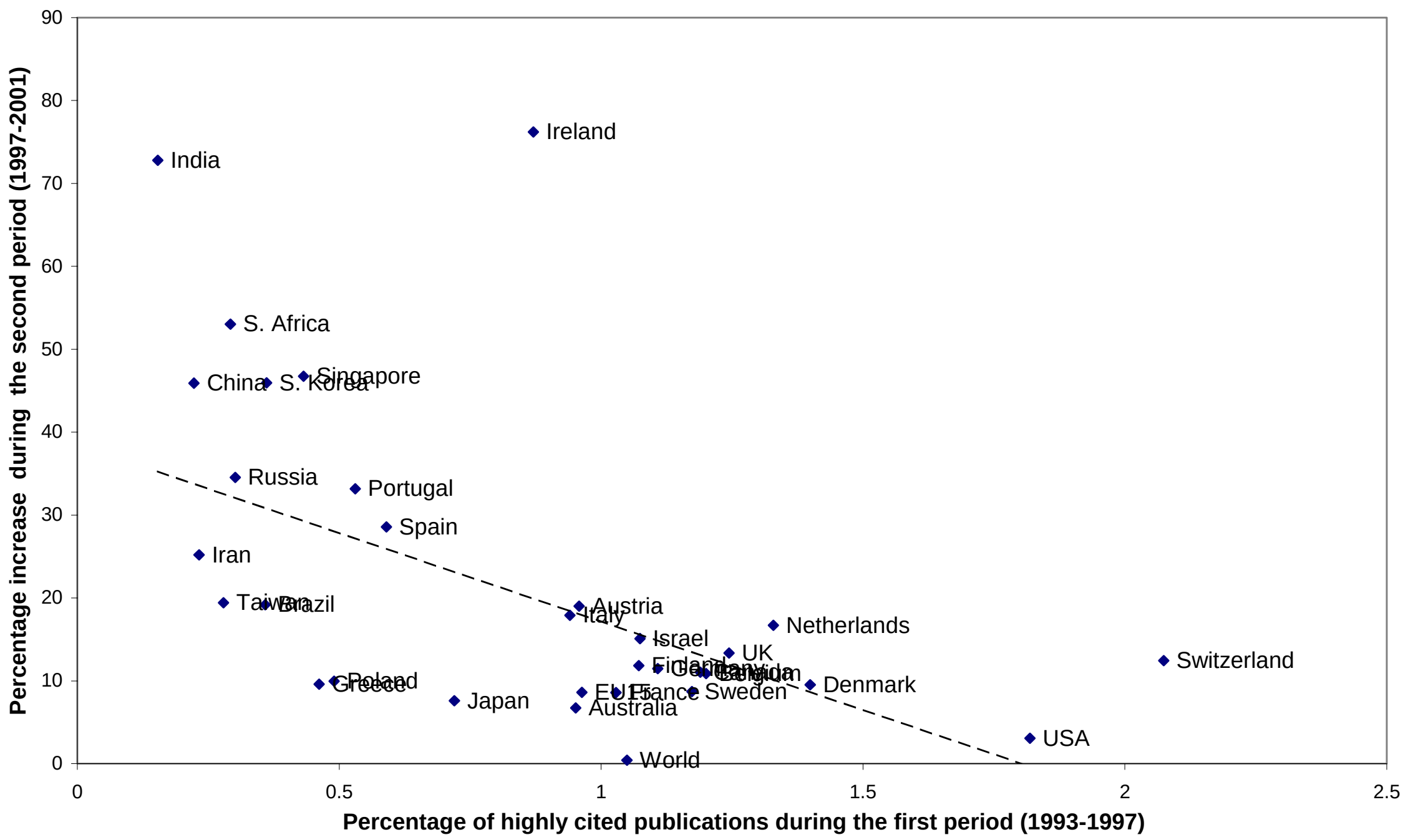

**Figure 5:** Growth of the number of most highly cited publications during the second period (1997-2001) normalized against previous contributions (1993-1997). (Source: King, 2004, at p. 312).

*3.2.2 Citation patterns of Chinese journals*

China is a large country not only in terms of its scientific publications, but also in the large number of scientific journals it produces. More than 4,400 of science and technology journals were published in China in 2001, and around half a million scientific papers are published annually in these journals (Jin & Rousseau, 2004). Two institutions have organized this domestic data in a format similar to that of the *Science Citation Index* of the Institute of Scientific Information in the USA. One database, the *China Scientific and Technical Papers and Citations Database (CSTPCD)*, is produced by the Institute of Scientific and Technical Information of China (ISTIC). In 2003, 1,576 journals were included in this database. The other, the *Chinese Science Citation Database (CSCD),* is produced by the Documentation and Information Centre of the Chinese Academy of Sciences, and covered 1,046 journals in 2001 (Leydesdorff & Jin, 2005).

The citation patterns of Chinese journals are different in their domestic and in their international environments. To explore these differences, we chose *Acta Chimica Sinica* as an entrance journal for two reasons. First, chemistry is among the fields in which China performs well (MOST & LCAS, 2004, at pp. 5-6 and p. 29). One can expect a more elaborate citation network for journals with a high profile than for journals in relatively weak fields. Secondly, this specific journal is covered both by the domestic and the international (*SCI*) databases, while it publishes in Chinese. For the domestic database, we used the *CSTPCD* as the data source.[2]

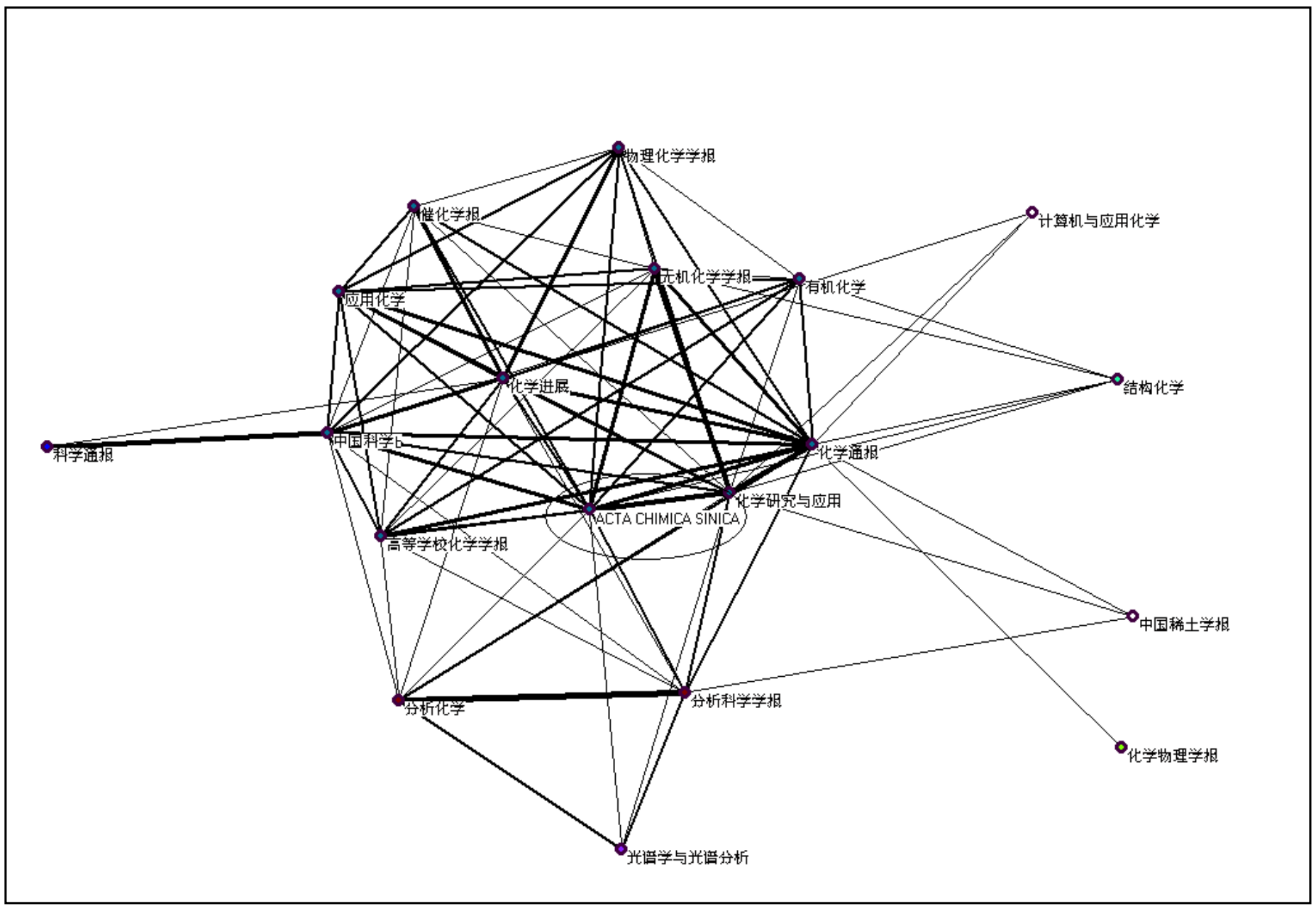

**Figure 6:** Cited pattern of *Acta Chimica Sinica* in the domestic database in 2002

Using *Acta Chimica Sinica* as the seed journal, Figure 6 shows that Chinese scientists cite publications in other domestic journals. Domestic journals in the same fields have close citation relations. In other words, they integrate with each other. This Figure 6 shows this for the being-cited patterns of the journals, but a similar picture emerges if we compare their citing behaviours. Chinese journals in this field (chemistry) are firmly integrated into a single unity in terms of their citation relations. The relation with marginal journals is focused on a few hubs, but the core group is extremely well interconnected.

How is the situation in the international environment? Figure 7 indicates that there are strong citation relations between *Acta Chimica Sinica* and international journals in the 'citing' dimension. This means that this journal is actively citing international journals

2 Jin & Leydesdorff (2004) provide a similar picture for this journal using the CSCD database in 2001.

in a pattern shared among these journals. In other words, Chinese scientists actively absorb knowledge produced by their international counterparts. International scientific literature has an impact on Chinese scientists.

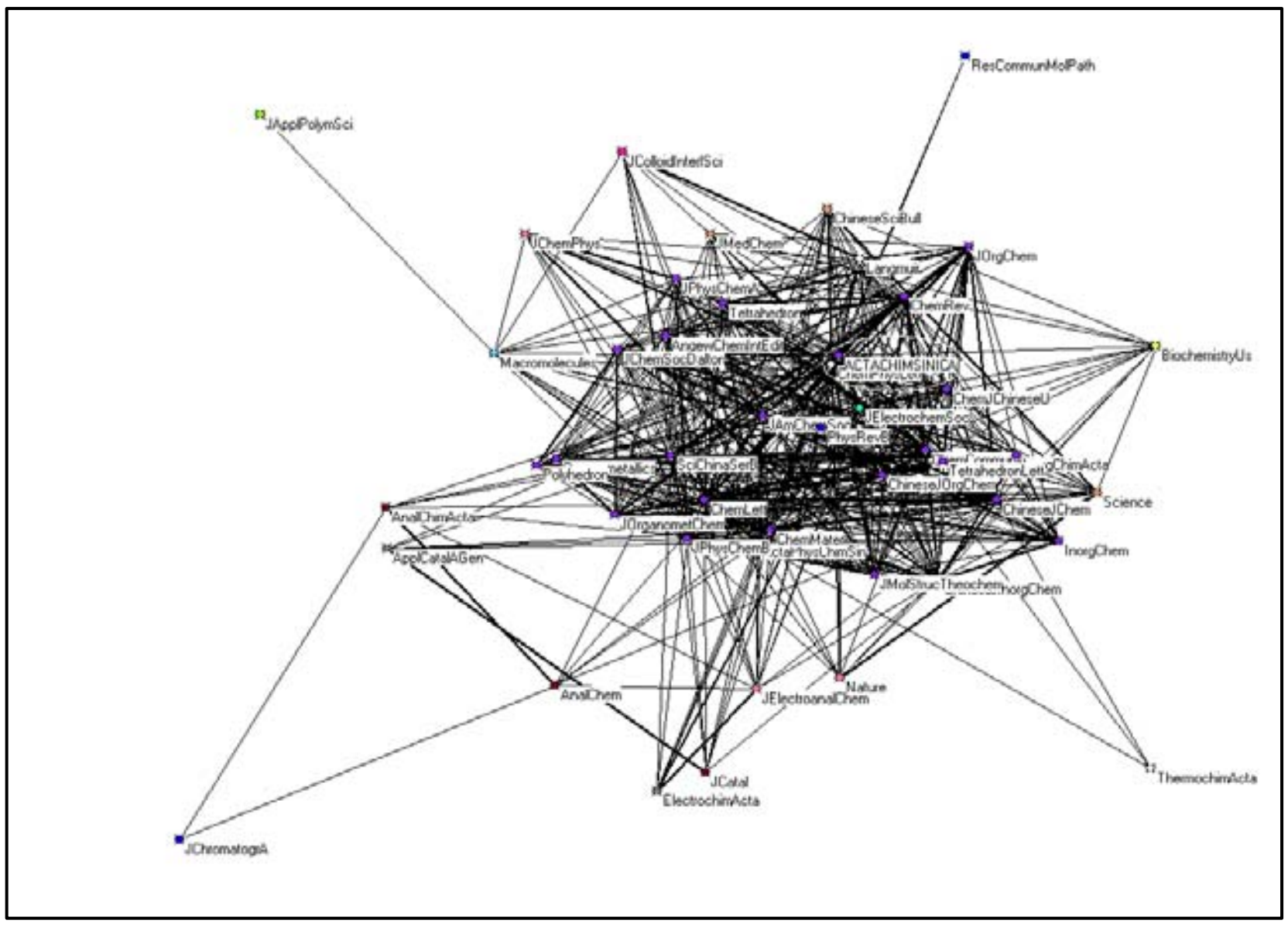

**Figure 7:** Citing pattern of *Acta Chimica Sinica* in the international environment (*SCI* 2002)

However, the 'cited' relations between Chinese journals like *Acta Chimica Sinica* and international ones are mediated by the *Chinese Journal of Chemistry*. In Figure 8, this journal functions as an articulation point between the Chinese and the international graphs. This means that the research output of Chinese scientists is not reflected by their international counterparts. Cited relations between international journals and Chinese journals are channelled through specific journals which function as citation windows on the Chinese literature. As *Acta Chimica Sinica* is published in Chinese, we conjecture that journals published in Chinese have not merged into the international academic environment even if they are included in the *Science Citation Index*. They form an isolated group within the citation relations of the set.

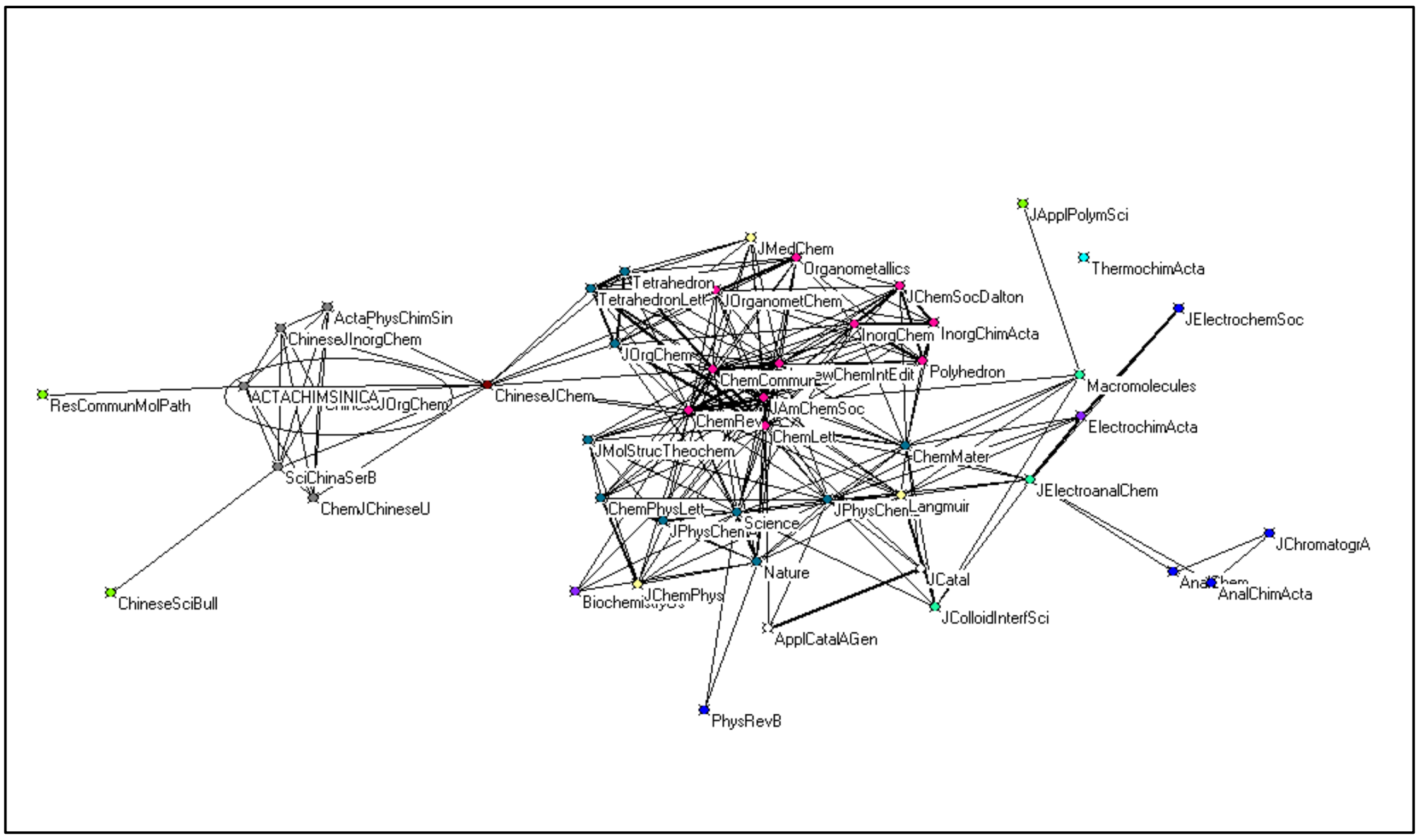

**Figure 8:** Cited pattern of *Acta Chimica Sinica* in the international environment (*SCI* 2002)

How is the situation of Chinese journals published in English? Let us check this by replicating the analysis for the *Chinese Journal of Chemistry* as an entrance journal. This journal functions as a bridge between international journals and the Chinese group of journals made visible in Figure 8 on the left side. Its cited situation is better than that of *Acta Chimica Sinica* (Figure 9).

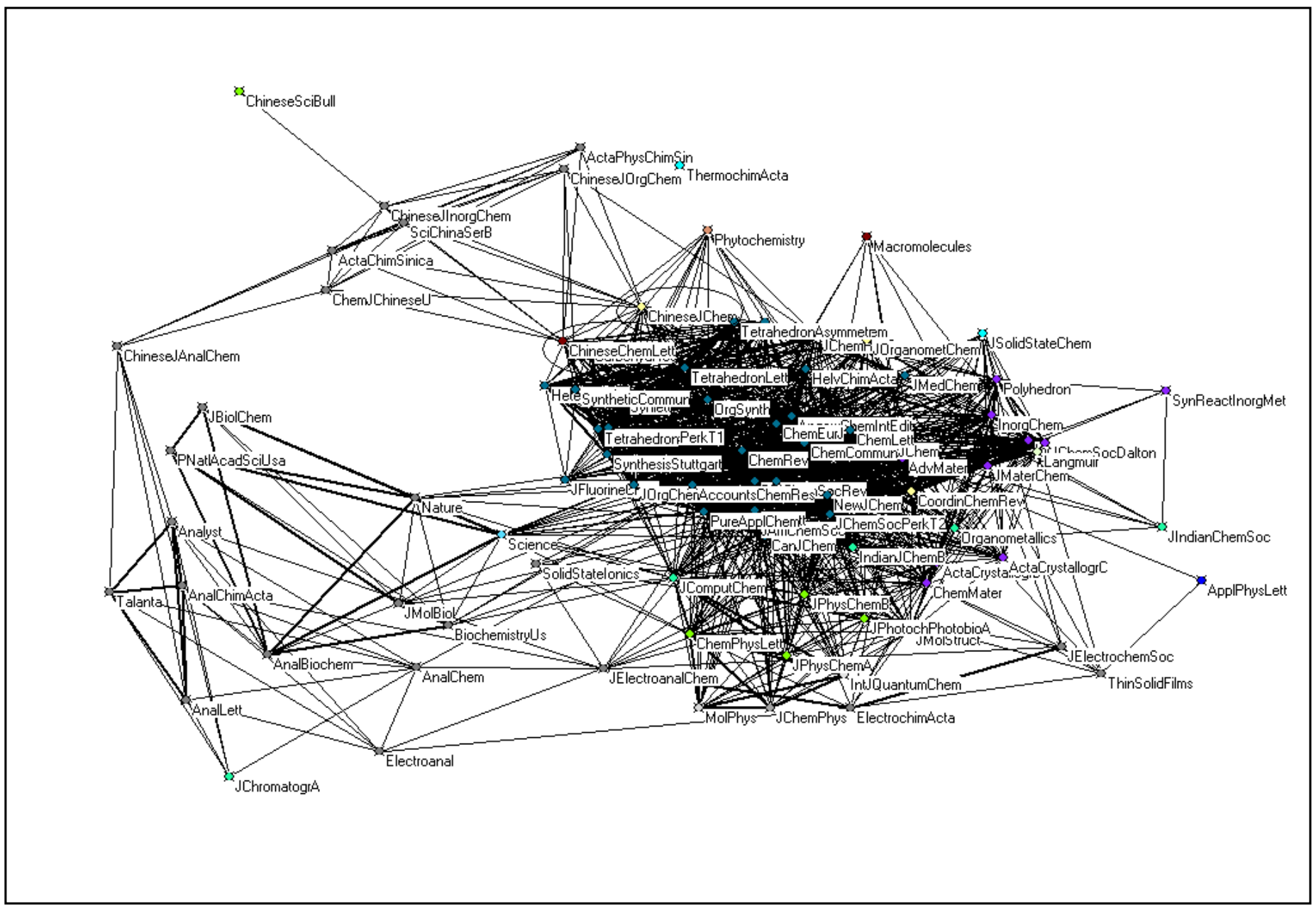

**Figure 9:** Cited relations of the *Chinese Journal of Chemistry* (SCI 2002)

The *Chinese Journal of Chemistry* is cited by its international counterparts, but its relations with other Chinese journals makes it special and therefore non-central. (A similar configuration can be shown for the other Chinese journal in English, that is, the *Chinese Chemical Letters*.) On the top left side of the picture are the journals that are published in Chinese, but included in the *Science Citation Index*. These results suggest that use of the Chinese language is an important factor that affects the international visibility of Chinese journals by isolating them from the international communication.

In summary, our analysis shows that Chinese journals are integrated with one another in the domestic citation environment. However, their citation patterns in the international environment are more complex. Inclusion in the *Science Citation Index* is not a sufficient condition for integration into the world system of scientific publications. Chinese journals are integrated with their international counterparts in terms of their citing relations, but the cited relations are not well established, especially for those published in Chinese. The language is a barrier for Chinese publications by authors who wish to be recognized in the international environment.

### 3.3 A focus on nanotechnology

Nanotechnology has been a key field for science and technology policies in recent years. In 2000, U.S. president Bill Clinton launched an initiative to promote nanotechnology entitled the *National Nanotechnology Initiative: Leading to the Next Industrial Revolution.* Since then, EU countries, China, Japan, and South Korea, etc., have all adopted nanotechnology as an S&T policy priority. The Chinese government, for example, declared nanotechnolgy a critical R&D priority in their *Guidance for National Development* in 2001. In the same year, the Chinese Ministry of Science and Technology, the National Development and Reform Commission, the Ministry of Education, the Chinese Academy of Sciences, and the National Natural Science Foundation jointly issued a *Compendium of National Nanotechnology Development (2001-2010)*. This can be considered as a strategic plan.

The potential importance of nanotechnology is generally acknowledged. Nanotechnology may have a significant influence on social and economic development and national security as well as people's daily lives. On the scientific side, developments in nanotechnology can be relevant for various fields such as physics, chemistry, material sciences, biology, and medicine (Meyer, 2001). In general, the development of new technologies provides challenges and new opportunities to existing fields of science (Rosenberg, 1982).

Since nanotechnology is an highly interdisciplinary field, it is difficult to identify which papers belong to this field. The set of *papers* with "nano" in their titles or keywords would also include those papers that satisfy this condition but do not really belong to the nanotechnology field. There may be some relabeling for opportunistic reasons (Studer & Chubin, 1980; Van den Daele *et al*., 1979). However, if we define the field in terms of *journals* with "nano" in their titles, we will certainly not be able to cover all papers in the nanotechnology field that are published in journals of related fields (for example, in physics, chemistry, materials sciences, etc.; cf. Bradford, 1934; Bensman & Wilder, 1998). Nevertheless, journals with "nano" in their titles may be a better source in terms of providing information in nano-papers (Schummer, 2004).

We experimented with various methods to delineate a journal set which would be representative of nanoscience and nanotechnology. The *Web of Science* in 2004 contained eight journals with “nano” in their title. Four of these journals (the *Journal of Nanoscience and Nanotechnology*, *Nano Letters, Nanotechnology,* and *IEEE Transactions on Nanotechnology*) have a high communality in their cited patterns given relevant journal environments. Factor analysis of the cited patterns can be used as an indicator for the development of new and emerging specialties. The journals with communality in their being cited patterns are recognized in the relevant environments as belonging to a single group (Leydesdorff *et al.* 1994).

To understand not only the current situation, but also historical developments in nanotechnology in terms of publications, we need data for at least three successive years. *IEEE Transactions on Nanotechnology* was not included in *SCI* before 2003, and can provide only two-year data. Therefore, we focused on the remaining three core nano journals, among which *Journal of Nanoscience and Nanotechnology* and *Nano Letters* were first covered by SCI in 2002, while *Nanotechnology* was covered as early as 1994.

Research in an interdisciplinary field like nanotechnology needs input from knowledge in other fields such as physics, chemistry, biology, or electronics. Publications cited by the core nano journals can be considered as relevant knowledge sources of nanotechnology. After some experimentation with different journal environments, we decided to consider all journals with citation relations above the one-percent level to the four *core* journals of nanotechnology as “nano-relevant” journals (Figure 10). These 85 journals cover the publication space for authors with communications potentially relevant for the nano-field. The authors who publish in these journals constitute also the human resources and the knowledge bases which can be activated by priority programs in nanoscience and nanotechnology.

Figure 10 shows that the three core *Journal of Nanoscience and Nanotechnology*, *Nano Letters,* and *Nanotechnology* indeed act as a core set at the interfaces among physical chemistry, material science and solid-state physics, while *IEEE Transactions on Nanotechnology* is more related to electronics and is less central to the interface. In other words, Figure 10 further legitimates the selection of these three journals (*Journal of Nanoscience and Nanotechnology*, *Nano Letters*, and *Nanotechnology*) as a core set in nanotechnology.

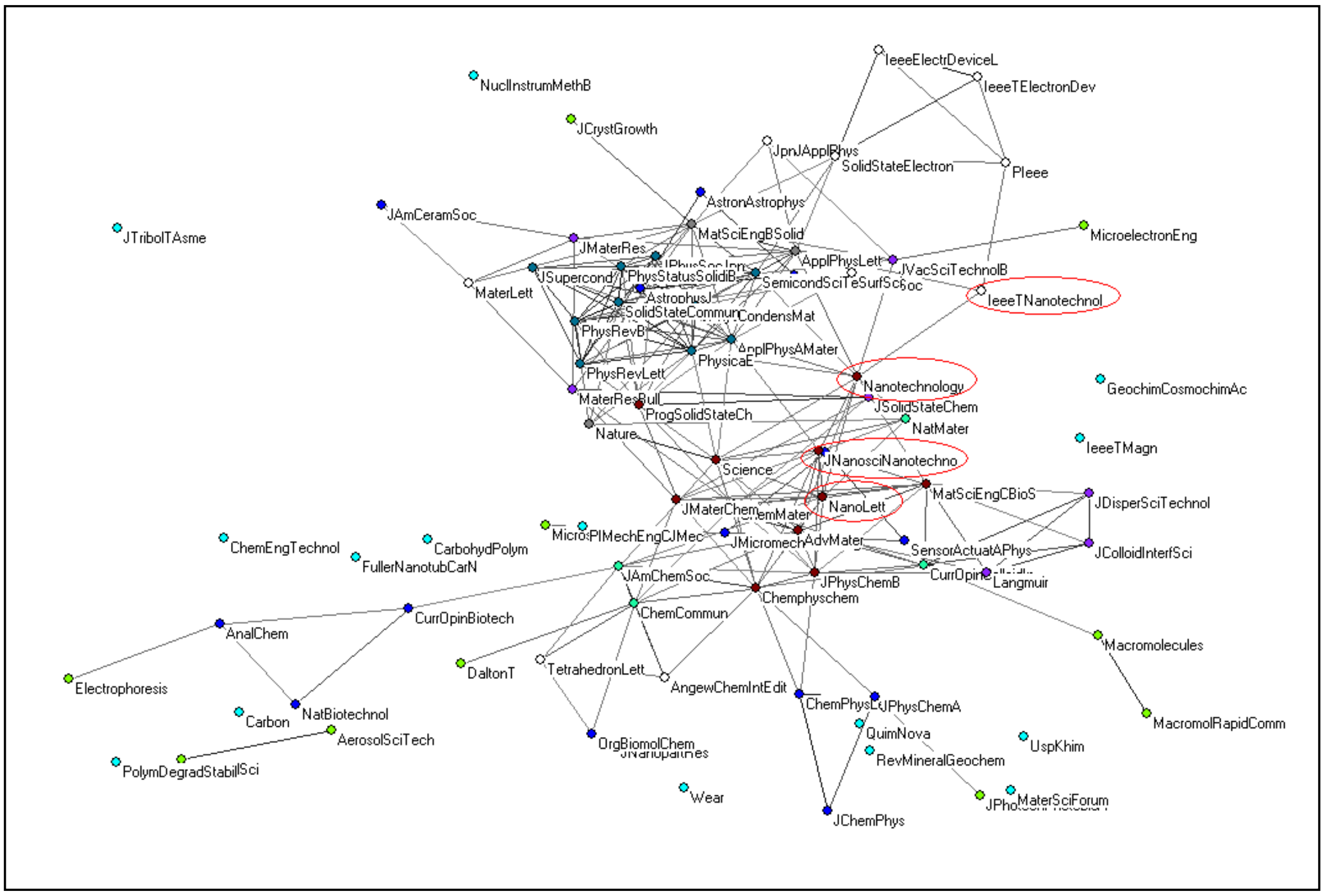

**Figure 10:** Cited environment of four core journals in nanotechnology and 85 "nano-relevant" journals (cosine ≥ 0.5).

We compared the national contributions in these two sets, that is, of the three core journals and the 85 nano-relevant journals. The mapping of the 85 nano-relevant journals is provided in Figure 10. The central position of the core journals is highlighted by using circles around each of them.

### *3.3.1 Results from core nanotechnology journals*

The year 2000, when the *National Nanotechnology Initiative* was published, can be considered as a watershed in the history of nanotechnology (President's Council, 2005). In addition to the implementation of national policies that assumed nanotechnology/nanoscience as a priority field, the influence was also reflected in the emergence of new journals in this field. While the journal *Nanotechnology* existed already before 2000, the other two core nano-journals (*Journal of Nanoscience and*

*Nanotechnology*, *Nano Letters*) were first published in 2001, and these journals were immediately covered by the *SCI* in 2002 (that is, after only one year of citations).

Since the first two journals (*Journal of Nanoscience and Nanotechnology*, *Nano Letters*) were not covered by the *SCI* until 2002, we collected data in two ways with the year 2002 as a dividing point. The first way was to collect the number of papers published in *Nanotechnology* from 1994 to 2004, in order to see the historical change; the second way was to collect the number of papers published in the *three core* journals distinguished above from 2002 to 2004. As before, we use only the articles, letters, notes, and reviews published in these journals as indicators, and compare China with the other major countries, the EU-15, and the EU-25 in terms of their research output on this indicator.

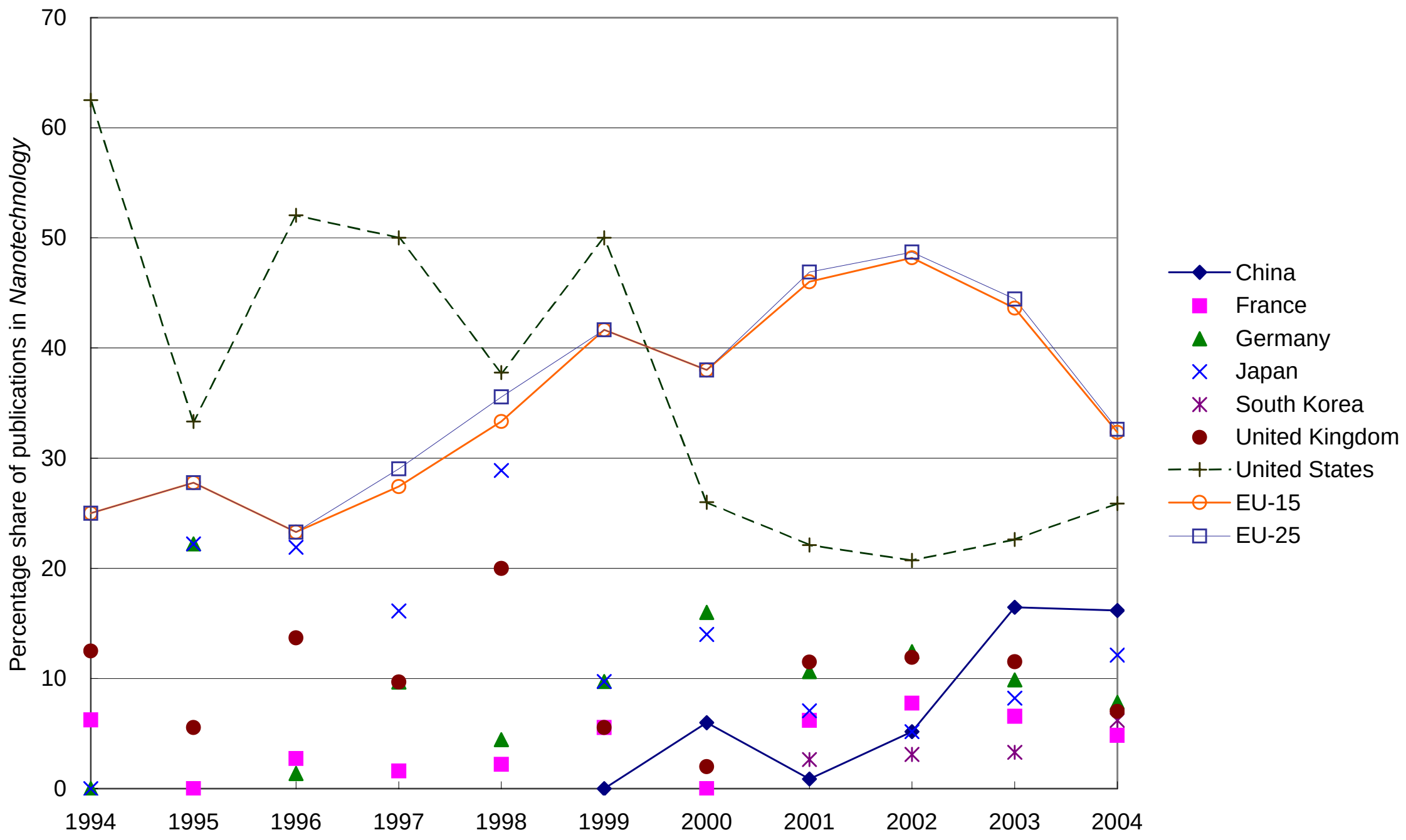

**Figure 11:** Share of publications in *Nanotechnology* (*Web-of-Science*, update date: 29 January 2005)

Figure 11 shows that the USA, the UK, and France were visible in this context from the very beginning, while research results with a Chinese address were published only since the year 2000. But China's progress is remarkable. Its world share increased obviously from 2001 to 2003, and it has become the second largest single country (after the USA) since 2003. The share of the EU-15 has surpassed that of the USA since 2000. The expansion of the EU with the ten accession countries does not make much difference on this indicator.

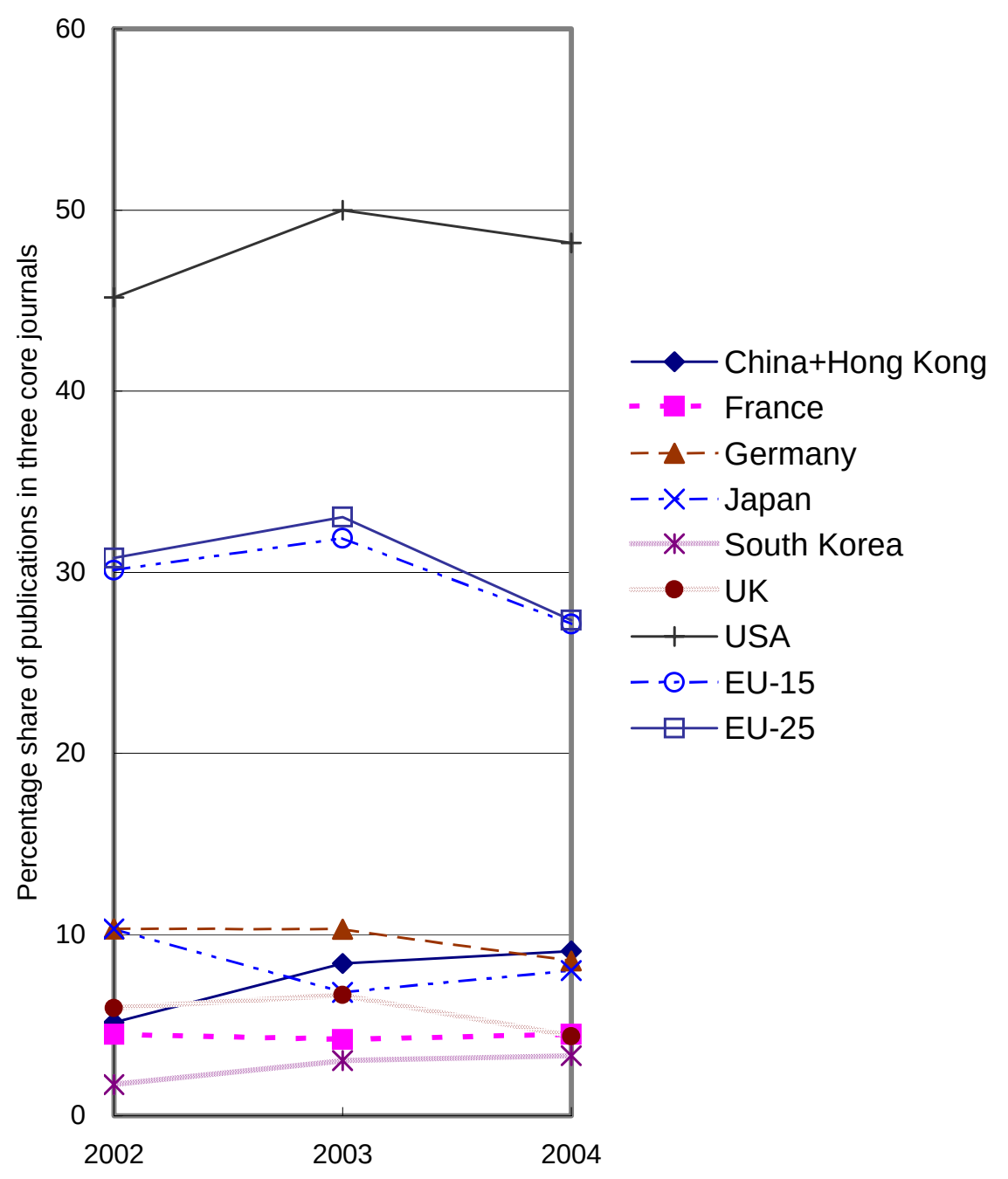

**Figure 12:** Share of publications in three core nano-journals (*Web-of-Science*, 29 January 2005).

Figure 12 expands the domain to the three core journals indicated above. One of the newly added journals (*Nano Letters*) is published by the American Chemical Society and thus part of an elite structure of journals highly integrated in the American system (Bensman & Wilder, 1998; Leydesdorff & Bensman, forthcoming). The USA, therefore, can expect to be represented in this set more than the EU. However, China has become the second largest single country in 2004 in terms of this indicator. While European countries have declined or remained stable, China and South Korea have maintained continuous growth trends in both sets.

*3.3.2 Results using the 85 "nano-relevant" journals*

Figure 13 shows the percentage of world share of publications in the 85 nano-relevant journals. The share of China increases in an exponential way, while the increase for South Korea is again linear. The pattern is very similar to the overall output patterns shown in Figure 2 above, but the percentages for some core countries are much higher than in the previous case. The UK has a higher percentage share when measured over the whole file; Japan and Germany have a larger world share in nano-relevant fields than in terms of their respective world shares of total publications.

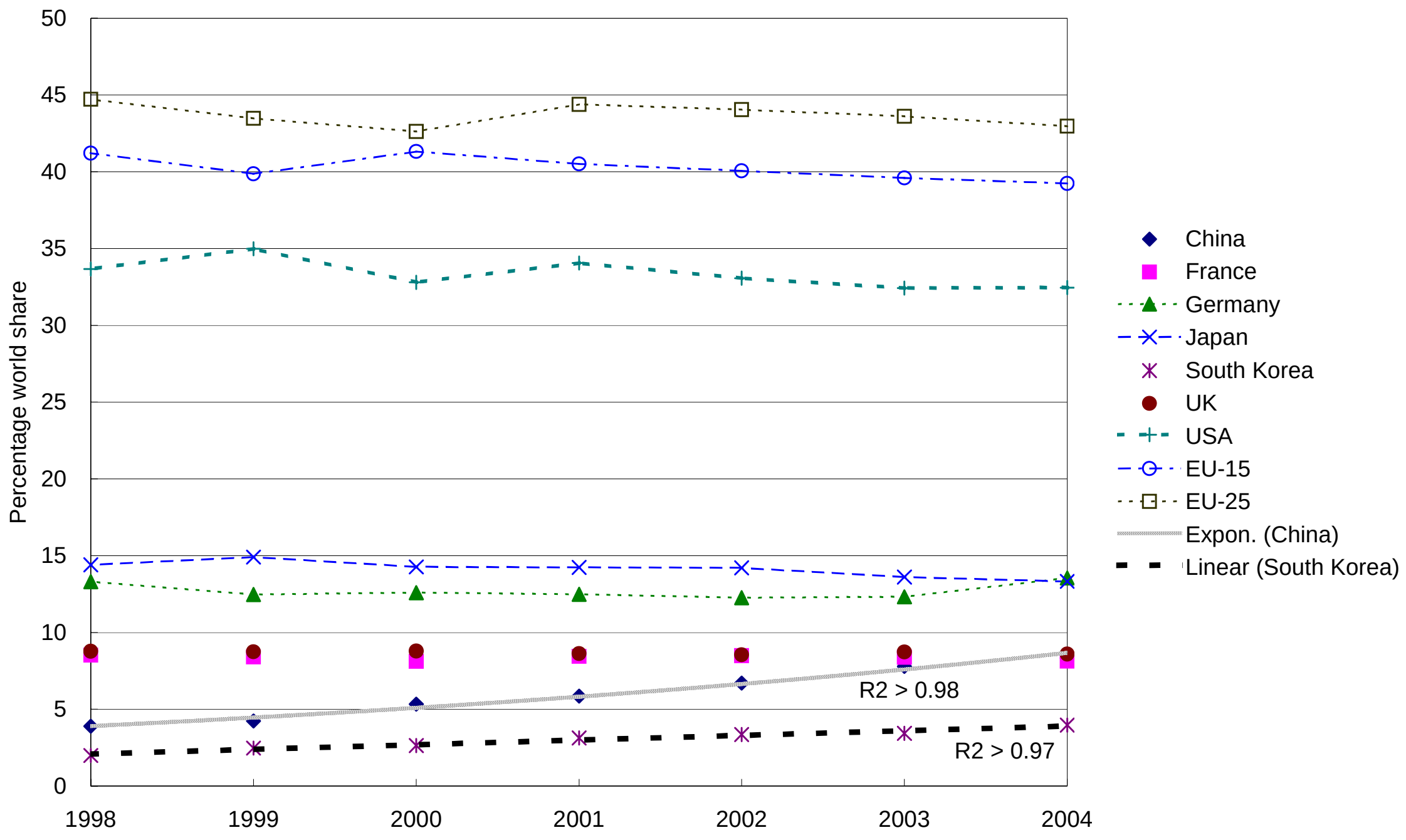

**Figure 13:** Percentage of world share of nano-relevant publications

For the USA, the percentage of world share of publications in nano-relevant fields is higher than the average for all fields of science as exhibited in Figure 2. After a decline in 2000, the percentage rises in 2001 after the publication of the *National Nanotechnology Initiative* in 2000. However, the USA's share decreased in later years since other relevant countries also began increasingly to invest in nanotechnology. The recent evaluation report the President's Council of Advisors on Science and Technology (2005, at p. 14) mentions this relative decrease on the basis of a much larger data set (based on collecting all documents using the keyword "nano*"). In the underlying study, Zucker & Darby (2005) found also China as the second largest producer of publications in this "nano*" area.

The EU-15 had a better performance compared to the USA in the nano-relevant set. Its world share has remained around 7% higher than that of the USA since 2002. However, the developing trend of both the EU-15 and the USA were similar, and they all lost some share after 2001. The addition of the other ten new EU countries contributed another 4% to the EU's world share. However, the trend of the EU-25 was similar to that of the EU-15 after 2001. The percentage of world shares of most of the other countries included in this study are also higher than their respective percentage shares of publications in Figure 2. These countries are Japan, Germany, France, South Korea and China.

In our opinion, these results indicate that some countries have surplus capacities to launch more research in nanotechnology, since expertise and manpower are available in nano-relevant sciences. One can expect more publications to come out of the countries with higher than average percentage world shares in nano-relevant fields because of available knowledge bases. However, China has continued additionally to increase its funding in this field as a priority area and, therefore, may be able to increase its share further provided that the conversion of input into output is efficient. Let us now turn to the issue of the relation between input and output.

## 4. Funding and Input/output ratios

The funding system of R&D in China is very different from that of Western countries where R&D is mainly conducted in universities. China's R&D is concentrated in public-sector research, partly because of the legacy of the Soviet-system of a strong Academy of Science.[3]

[3] There is still some debate about purchasing power parity (PPP) in relation to the Chinese RMB's exchange rate (Davies, 2003; Shi, 2004). Furthermore, the government and higher education expenditures cover all fields of natural sciences (including agricultural and medical sciences) and engineering (NSE), as well as social sciences and humanities, while the business enterprise sector covers only the fields of NSE. There are only a few organizations in the private non-profit sector. Hence, no R&D survey has been carried out in this sector, and consequently this data is not available. However, the line between public and private sectors in China is not easy to draw, due to this country's public "branch institutes." In the past, research by these institutes was completely funded by the government. With the further reforms of the S&T system many of these institutes have been transformed into corporations, and no longer receive public funding. Since the system is still in transition, some institutes receive both public and private funding.

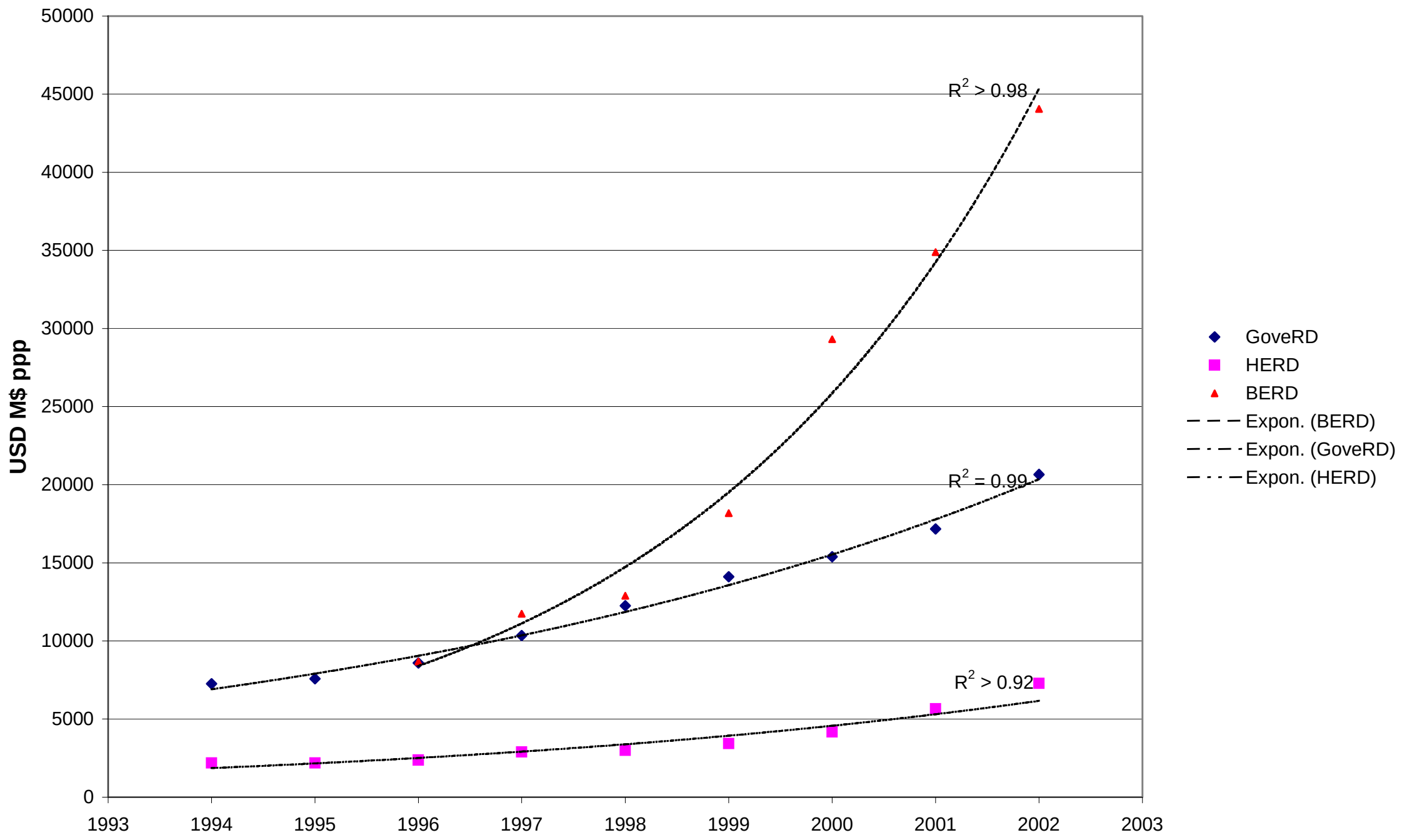

**Figure 14:** China's funding of R&D from 1994 to 2002.

Figure 14 shows that the Chinese government has effectively used its public-sector research potential to boost the knowledge-based economy of the country. This investment has triggered business expenditure in R&D (BERD) during the second half of the 1990s. All growth curves are exponential, but the rates of growth in GOVERD ("Government Intramural Expenditure" or, in other words, public-sector research) are approximately three times as large as in the university sector, and the funding levels are also almost three times as large. This pattern of government spending in R&D in China is different from that of the West. BERD surpassed GOVERD in 1997, and the distance between the two has increased ever more since then.

The sustained increase in R&D funding in China is, of course, backed by the rapid growth of the country . The ratio of GERD/GDP is an important indicator. The "Lisbon agreement" (European Commission, 2000), for example, set 3% GERD/GDP as a target for EU nations to be reached by the year 2010. In order to obtain a picture of the relations between GERD and GDP, we gathered related data from OECD *Main Science and Technology Statistics,* and included data of the above-mentioned countries, EU-15 and EU-25 (Figure 15).

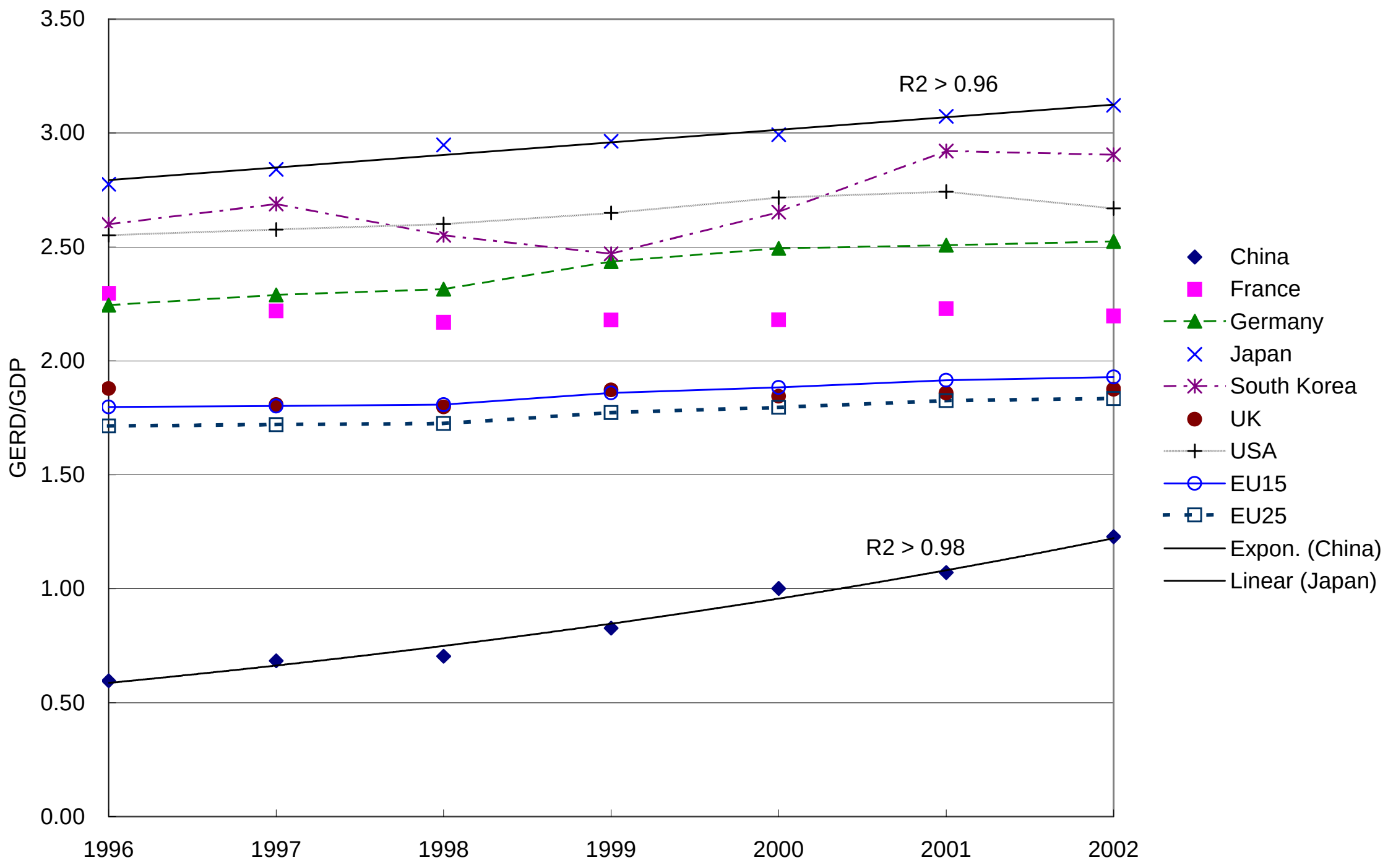

**Figure 15:** GERD/GDP of the major countries, the EU-15, and the EU-25 (OECD, *Main Science and Technology Statistics*, 2005)

Despite the enormous increase of the GDP in the nominator, the Chinese GERD/GDP ratio has been increasing exponentially since 1996. From 1996 to 2002, this figure has risen from 0.60% to 1.23% (OECD, 2005). The percentage in China has doubled within only seven years, while this percentage has remained relatively stable (approximately 1.9%) for the EU-15 (OECD, 2005). Among the other major countries, Japan has had the highest GERD/GDP, and it has been able to maintain a linear growth for this indicator. Since 2001, Japan is the only country whose GERD/GDP has surpassed 3%. South Korea's GERD/GDP in 2001 and 2002 were 2.92% and 2.91% respectively, but among the European nations only Sweden and Finland are above the 3% level.[4] Among the other EU countries, the GERD/GDP ratios of Germany and France were higher than that of the EU-15 on average, while the UK fluctuated around the average in terms of this indicator. Interestingly, given the Lisbon objective, one does not yet see convergence among EU nations on this indicator.

[4] Finland had a GERD/GDP ratio of 3.46% in 2002, and Sweden 4.27% in 2001 (OECD, 2004).

Industry does not primarily publish scientific articles, but industrial innovations are reflected in patents (Jaffe & Trajtenberg, 2002). However, we focus in this study on the scientific side of the development. From this perspective, the institutional patterns in Chinese output are more in line with the western counterparts of China. For example, Figures 16a and 16b show the shares of publications of universities, research institutes, hospitals, and business included in 2003 in the domestic *CSTPCD* and the international *SCI*, respectively (ISTIC, 2004). As elsewhere, universities are the largest shareholders, while hospitals and research institutes also make a considerable contribution to Chinese S&T publications. Hospitals are important producers of publications in the domestic sciences, but not internationally. Public research institutes made in absolute numbers the second largest contribution to international science. Enterprise research contributes only marginally to international publications, but it provides a share of 6% of the national publications.

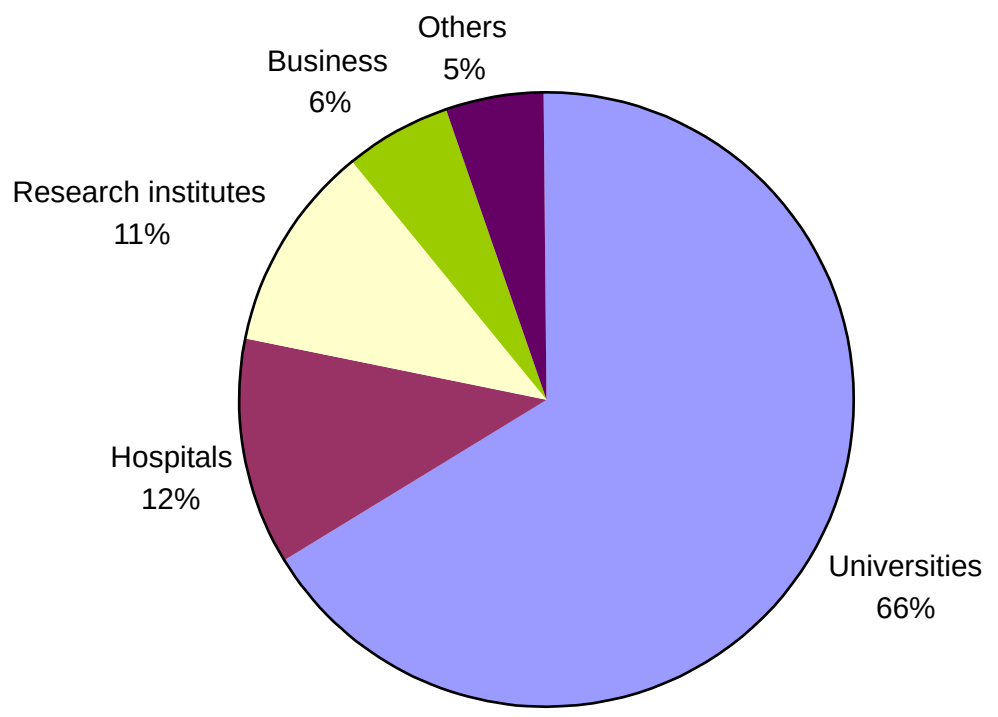

**Figure 16a:** Distributions of Chinese domestic publications in 2003 (CSTPC; ISTIC, 2004)

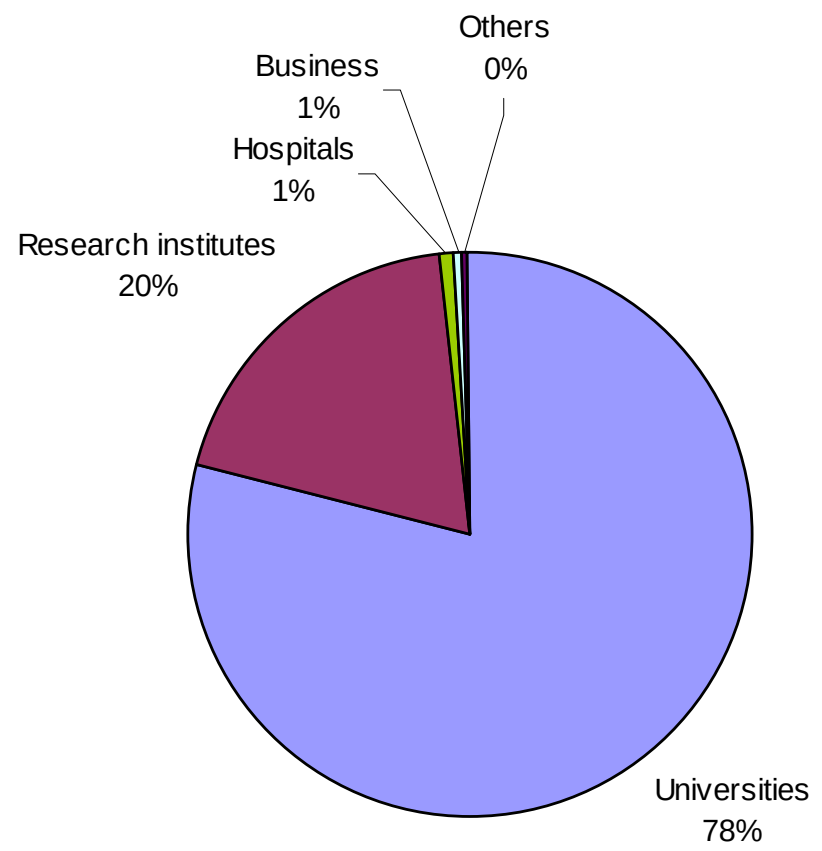

**Figure 16b:** Distributions of Chinese international publications in 2003 (*SCI expanded*; source: ISTIC, 2004)

In order to make this data comparable with those of western nations, we have added the Government Intramural Expenditure on R&D (GOVERD) and Higher Education Expenditure on R&D (HERD) together as the total government input to R&D. These normalized expenditures (OECD, 2004) can than be plotted against output measured in terms of the percentages of world share of publications as provided in Figure 2 above. Figure 17 shows the relations between input and output for the countries under study and the EU.

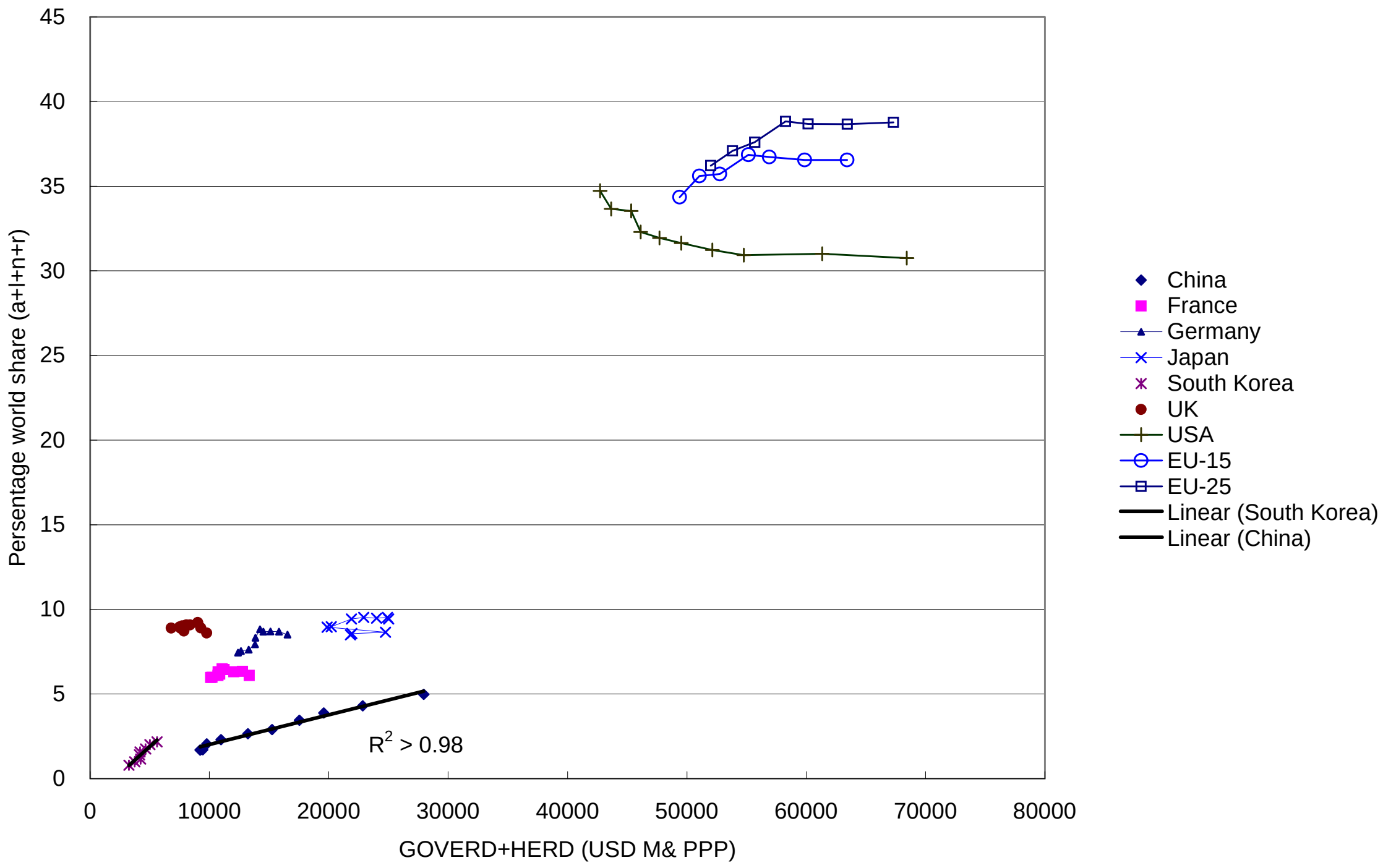

**Figure 17:** Input/output of major countries, EU-15, and EU-25

China's output shows a linear relation with input ($r^2 > 0.98$). In other words, the increase in R&D expenditure by the Chinese government is used efficiently: with the increase of R&D investment, the country's percentage of world share grows proportionally. The dynamics can thus be considered as a self-reinforcing mechanism. The possibility to publish internationally and to participate in the knowledge-based economy is continuously reinforced within the Chinese system. For example, researchers at many universities receive a considerable bonus in their salaries when they publish in journals that are included in the *Science Citation Index*.

The link between funding and scientific production is not deterministic, and probably even less so for highly developed countries. In the Western (liberal) model, scientific development is relatively autonomous, while the Chinese government probably has more steering mechanisms, since it has also inherited elements of the old Soviet model. Notwithstanding these differences in the mediating mechanisms in the various countries, Figure 17 shows several interesting features at the systems level. For example, German unification has led to a stepwise increase of output in relation to similar input in the middle of the 1990s. Japan exhibits a different pattern because this country first decreased funding and then expanded it again in the later part of the 1990s. The decrease did not lead to a loss in the percentage of world share, but the reversal of this trend has made the system a bit more productive.

The USA greatly increased intramural funding of R&D within government agencies after "9/11," but this increase has not been reflected in an increase of world share of publications. One of the reasons might be that the emergence of other scientific countries like China and South Korea has put pressure on traditional advantages. However, several European countries and Japan have also improved their performance on these indicators, albeit more modestly. Another reason might be that a large proportion of this American funding is spent on classified research which does not lead to publications. The input levels between the EU and the USA were similar, but the efficiency in the EU was higher than that of the USA. In other words, during the 1990s the EU has become more productive in terms of scientific publications than the USA.[5]

The development of the input-output ratio for South Korea is most remarkable because linear growth was maintained during a number of years. As noted, Korea has been a member of the OECD since 1996 and it has adopted a western pattern of funding. Perhaps more than any other OECD-country, however, South Korea defines its performance also in relation to China as a major competitor. Although South Korea has not been able to keep pace with China in extending its absolute data of funding, its GERD/GDP is still much higher than that of China: in 2002, the GERD/GDP of South Korea (2.91%) was more than twice that of China (1.23%).

## 5. Discussion

We have mainly relied on three databases: the *Science Citation Index* of the ISI, the *China Scientific and Technical Papers and Citations Database* of the ISTIC, and the *Main Science & Technology Indicators* of the OECD. These databases are statistical and therefore introduce uncertainties and potential sources of error. We noted the ongoing debate in the literature about the conversion of the Chinese currency into normalized US dollars in the case of the OECD database. The two publication and citation databases (that is, the domestic one and the international one) both have a problem of representation with reference to the underlying population. In the case of the People's Republic, we happen to have a precise count of 4,497 scientific journals published in 2003 (Ren, 2005). Of these 1,506 were included in the *CSTPCD* and 67 in the *SCI*.

[5] There is a parlance that authors might increasingly split one paper into two or even three, which would cause "paper inflation." However, there is no systematic evidence for this inflation, and there is no *a priori* reason why this tendency would be different among nations.

Inclusion in the *SCI* has been debated in terms of national, language, and disciplinary biases. Van Leeuwen *et al*. (2001), for example, have argued that the language bias of the coverage has consequences for international comparisons of national research performance. Sivertsen (2003), however, found no bias of the ISI-database when evaluating Scandinavian publications. The ISI has admitted a bias against including journals published in languages other than those using the Latin alphabet.

For the purpose of our study, the language bias does not pose an analytical problem given our research question. The focus of this study is on the visibility and the translation of the Chinese S&T capacity into the international arena. The latter is partly operationalized as publications in English. Publications of Chinese authors in languages other than Chinese and English (e.g., in Russian) have not been considered. Given this research question, we found also a representation of journals in Chinese within the international database, and although this can obviously be considered as an underrepresentation of the Chinese potential, we were able to signal the danger of the relative isolation of publications in the Chinese language despite their inclusion in the ISI database. Of course, the Chinese percentage world share of publications and citations in this database would increase further if more Chinese journals were included, but for the purpose of our research the possible underrepresentation only strengthens our conclusions about the emergence of China as a leading nation in science.

The disciplinary bias of the ISI-database in favour of biomedicine and the life sciences is a more important reason for concern. Recently, Park *et al*. (2005) signaled that the research portfolio of the Netherlands is much more compatible with the journal portfolio of the *SCI* than with that of South Korea. The authors conclude that this might have a considerable effect when comparing these two countries using this database. Asian countries like China and Korea have strengths and weaknesses in the portfolio that are different from those of Western countries (and the latter group is not homogenous as well). This effect may partly explain the enhanced visibility of China in a subset like the one which we constructed as "nano-relevant."

As noted, the "nano core" journals are more deeply integrated into the elite journal structure of the United States. These journals can be expected to have a bias in favour of accepting papers from elite institutions in the USA and in other advanced countries, and may not be easy to access by scholars from more peripheral locations. We noted that the Chinese contribution in these core journals has been increasing considerably, as has the Chinese contribution to the one percent most highly cited papers.

Part of this may be the result of international collaboration with Chinese authors. According to the ISTIC (2002, 2003), the number of internationally coauthored publications with at least one Chinese address increased from 7,807 in 2002 to 11,739 in 2003. Internationally coauthored papers in 2003 thus accounted for 23.6% (that is, more than one-fifth) of its total publications included in the *SCI*. (Among the 11,739 publications, 5,942 [50.6%] papers were first authored by authors with a Chinese address.) The first five countries to cooperate with China are the USA, Japan, Germany, the UK, and Australia. Four of these countries rank as the first four in terms of publications in 2004. The partnership with Australia points to the importance of the geographical factor.

When coauthorship relations are normalized, China appears to have become well integrated into the Asian-Pacific region during the 1990s (Wagner & Leydesdorff, 2005). Some authors have recommended normalizing S&T indicators in terms of the size of the respective populations. In the case of China as a developing nation, such a *pro capita* normalization would have dramatic effects, and the phenomena to which we wished to draw attention would completely disappear. The size of the *scientific* community in a nation could be considered as another factor. A large scientific community may lead to a relatively large within-country citation rate, while scholars in small nations may have to rely more on international colleagues. A correction of the citation rates of the USA or China for the within-country citations, however, would have very large effects on the citation indicator (Seglen, 1997). This indicator would have a meaning in terms of the networking of international collaboration and influence more than in terms of national performance.

**6. Conclusions**

China has become a major player in the production of scientific papers. Its contribution to world science shows exponential growth (Figures 1 and 2), which is unique in the world. Scientific research activities in the EU countries are the most productive in the world. In terms of this indicator, the EU countries have left the USA behind during the 1990s. More than 68% of the scientific papers included in the *SCI* are from the EU countries and the USA together. In other words, these countries make the biggest contribution to world science, while the contribution of Asian countries, mainly Japan, China, and South Korea, are in second place (Figure 2 and Table 1).

Along with the exponential increase of scientific publications, the citation rates of Chinese publications are increasing exponentially as well (Figure 3). Other indicators measuring the impact of publications, such as the percentage of world share of citations and the number of most highly cited publications, also demonstrate the increased status of Chinese publications (Figures 4 and 5). Chinese journals play important roles in the communication of Chinese scientists in the domestic environment. However, in the international environment, Chinese journals have integrated with their international counterparts in terms of Chinese papers citing articles written by others, but the cited relations have not yet been established. Journals which publish in Chinese are not often cited in the international literature even if they are included in the *Science Citation Index*.

China's performance in nanoscience and nanotechnology is remarkable as well. Although it started research in this field later than the other major countries like the USA, France, Germany, and Japan, China's world share of publications in nanotechnology has increased rapidly. Using various indicators, we found China in 2004 in the second position behind the USA. China's potential in further expanding research in this field is large as well, which can be seen through its higher world share of publications in nano-relvant fields compared with its world share of publications over the entire file. In 2004, China's world share of publications was on average 6.52%, while its world share in nano-relevant publications was 8.34%. The increase rates for the exponential fits of the curves are correspondingly higher (the coefficients of the exponents are 0.126 and 0.133, respectively). Again, more than half of the world publications in nanoscience and nanotechnology are from the USA and the EU countries, while China, Japan, and South Korea account for most of the remaining share (Figures 11 and 12).

The Chinese government pays unprecedented attention to the development of science and technology and the transition to a knowledge-based economy. More than any other country in the world, funding for R&D is growing not only absolutely, but also relative to the spectacular growth in the gross national product (Figures 14 and 15). It is noteworthy that business investment in R&D is increasing even faster than government expenditure. This increased investment has been reflected in the output: China's output in terms of scientific publications is also increasing exponentially (Figures 2 and 17), and thus one may assume an efficient coupling between input and output. Among the countries studied, Japan has the highest GERD/GDP ratio, and its trend shows linear growth, but like most countries it spends more to maintain approximately the same share (Cozzens *et al.*, 1990). The EU countries grow slowly in terms of this indicator, but they are still more than one percent below the target of 3% (GERD/GDP) set by the "Lisbon" agreement. The USA's investment is higher than that of the EU countries, but with less output in scientific publications.

As mentioned above, there are 4,497 scientific journals published in mainland China in 2003 (Ren, 2005), among which, only 67 were included in SCI in that year, while 274,438 papers were included in the 1,576 journals covered by the CSTPCD in 2003. This data shows that China has huge human resources in science and technology. With exponentially increasing funding and proper guidance of scientific policies, one can expect that China's increased momentum in scientific publications is sustainable. The problem for China is not the production of scientific publications, but the dissemination and visibility of scientific publications in the international communication system of science.

## 7. Policy implications

During the period 1997-2001, China's percentage of world share of citations was only 1.56%, standing in the thirteenth position in the world. It seems that Chinese citation rates are not compatible with its number of publications. The c/p (citations over publications) ratio was only 2.96 during this period, while the highest ratio was that of 9.69 in Switzerland (King, 2004). Some authors have analysed this lagging behind of citations as an indicator of a lack of quality in the system (Jin & Rousseau, 2004; Poo, 2004; Wu, 2004). According to these authors, the following factors in the production system of science would have a negative effect on the impact of Chinese science: the institutional evaluation system for research proposals; research output is rigid; investment in basic research is too low; and the higher-education system is not sufficiently internationalized. However, our analysis points in another direction. The efficiency of the Chinese system is not low and is still increasing. The citation rate is also increasing exponentially, but this development is delayed and not proportionate to the increase in the publication rate.

In addition to continued attention to improving the quality of research, it therefore may be urgent to take measures to increase the visibility of Chinese publications. Most scientists are inclined to think that a high-quality research result will be noted and resonate in the international communication without further efforts. Of course, a high-quality paper needs to contain novelty, but this is not a sufficient condition for its diffusion. The production and diffusion of knowledge are different dynamics. The dynamics of production involve developing and designing a research project, gathering funding and other resources, and then conducting the research. It involves communication between researchers and research administrators who control research funds.

The diffusion rate is limited by the quality of the communication dynamics among the authors, their audiences, and the editorial boards of scientific journals. At this stage, the quality of communication skills (e.g., one's ability to organize a paper in English), communication channels, and communication networks are key factors that affect the communication results and therewith the visibility of publications. From this second perspective, we are able to make the following suggestions for improving China's performance at the global level of science:

✧ *Focus on international journals*

High-quality journals attract more scientific readers, and publication in such journals leads to higher visibility and therefore higher citation rates. Chinese scientists may consider changing their focus from domestic journals to international ones, especially those with high impact factors. When analysing the citation patterns of Chinese journals, we found that Chinese journals have remained isolated from the international community even if they are published in English and included in the *SCI*. The visibility of Chinese authors in world science has remained relatively low. This means that papers published in Chinese journals cannot have as much impact as those published in international journals in English. Therefore, publishing papers in international journals is one way to increase the international visibility of Chinese papers.

China has a communication system of national journals which may be extremely important for the diffusion of scientific knowledge into the economy (Leydesdorff & Jin, 2005). This system is a legacy of the previous period with its emphasis on autarchy. In the current transition to a knowledge-based economy, this "Mode 2" set of journals that is already integrated with its applicational contexts (Gibbons *et al.*, 1994) may itself be an asset. However, these journals should not be used as alternative output channels for academic publications which can also be published internationally. The differentiation between these two publication systems can be made a subject of further reflection.

Competition to publish papers in international journals, especially high impact journals, is fierce. With the pre-condition that research is original and/or leading-edge, the author's writing ability in English is very important in this stage. A scientist needs to make his/her paper well-organized and to highlight the points that are original or creative. In general, the ability to present a paper that properly reflects the significance of research is a very important skill in scientific communication.

✧ *Improve the quality and visibility of Chinese journals*

Chinese journals are still by far the main channels for Chinese scientists to communicate with their international counterparts. Their quality and visibility directly influence the citation rates of Chinese papers negatively. High-quality journals need high-quality papers. Absorbing high-quality papers, especially papers produced by influential world scientists, affects a journal's quality. The relatively low quality and international attractiveness of Chinese journals has already been recognized by the Chinese government. The Ministry of Science and Technology (MOST) plans to select some promising Chinese journals and help them raise their quality to become world-class S&T journals, by means of financial support (SciDev. Net, 2004). Other journals may consider to publish online and as open-access journals in order to enhance their visibility (Harnad, 2004).

✧ *Strengthen international collaboration*

Our study shows that the EU countries and the USA are the major contributors to world science. In other words, the majority of world-class scientists are located in these countries, and these countries have excellent research conditions (laboratories and funds) as well. In addition to helping Chinese scientists improve their research skills and expand their perspectives, cooperation with scientists in these countries can expand the communication networks available to Chinese scientists and provide access to research networks for Chinese publications, thus raising their visibility. After a cooperation project is finished, Chinese scientists should remain in close touch with their international counterparts to ensure that collaboration and communication are sustained. In recent years, China has broadly expanded its scientific cooperation, which can be seen through the continuously increased number of co-authored scientific papers (Jin & Rousseau, 2005; Wagner & Leydesdorff, 2005). However, more chances and possibilities still exist and remain to be explored.

**Acknowledgement**

We would like to thank Ma Zheng of the statistics team of the Institute of Scientific and Technical Information of China for providing us with relevant data and we are grateful to Wu Yishan and Ronald Rousseau for useful comments on previous versions of this paper.